\documentclass[conference]{IEEEtran}
\IEEEoverridecommandlockouts
\usepackage{cite}
\usepackage{amsmath,amssymb,color,graphicx,caption,subcaption,comment,tikz,url,latexsym} 
\usepackage{algorithmic}
\usepackage{graphicx}
\usepackage{textcomp}
\usepackage{pgfplots}
\usepackage{xcolor}
\usepackage{url}
\usepackage{multirow}
\usepackage{mathtools}
\usepgfplotslibrary{units}
\usetikzlibrary{spy,backgrounds}
\usepackage{pgfplotstable}

\def\BibTeX{{\rm B\kern-.05em{\sc i\kern-.025em b}\kern-.08em
		T\kern-.1667em\lower.7ex\hbox{E}\kern-.125emX}}

\newtheorem{thm}{Theorem}

\newtheorem{definition}{Definition}

\newtheorem{remark}{Remark}
\usepackage{bbm}
\newcommand{\M}{\mathsf{M}}
\newcommand{\m}{\mathsf{m}}
\newcommand{\len}{\textrm{len}}

\newcommand{\fix}{\textnormal{max}}
\newcommand{\Tx}{{{T$_{X}$}}}
\newcommand{\Rel}{{{R$_{Y}$}}}
\newcommand{\Rec}{{{R$_{Z}$}}}
\newcommand{\E}{\mathbb{E}}

\begin{document}
	\interdisplaylinepenalty=0
	\title{Two-Hop Network with Multiple Decision Centers under Expected-Rate Constraints\\
	}
	\author{\IEEEauthorblockN{Mustapha Hamad}
		\IEEEauthorblockA{\textit{LTCI, Telecom Paris, IP Paris} \\
			91120 Palaiseau, France\\
			mustapha.hamad@telecom-paris.fr}
		\and
		\IEEEauthorblockN{Mich\`ele Wigger}
		\IEEEauthorblockA{\textit{LTCI, Telecom Paris, IP Paris} \\
			91120 Palaiseau, France\\
			michele.wigger@telecom-paris.fr}
		\and
		\IEEEauthorblockN{Mireille Sarkiss}
		\IEEEauthorblockA{\textit{SAMOVAR, Telecom SudParis, IP Paris} \\
			91011 Evry, France\\
			mireille.sarkiss@telecom-sudparis.eu}
	}
	\allowdisplaybreaks[4]
	\sloppy
	\maketitle

	\begin{abstract}
	The paper studies distributed binary hypothesis testing over a  two-hop relay network where both the relay and the receiver decide on the hypothesis. Both communication links are subject to \emph{expected} rate constraints, which differs from the classical assumption of maximum rate constraints. We exactly characterize the set of type-II error exponent pairs at the relay and the receiver when both type-I error probabilities are constrained by the same value $\epsilon>0$. No tradeoff is observed between the two exponents, i.e., one can simultaneously attain maximum type-II error exponents both at the relay and at the receiver. 
For $\epsilon_1 \neq \epsilon_2$, we present an achievable exponents region, which we obtain   with a scheme that applies different versions of a basic two-hop scheme that  is optimal under \emph{maximum} rate constraints. We use the basic two-hop scheme with two choices of parameters and rates, depending on the transmitter's observed sequence. For $\epsilon_1=\epsilon_2$, a single choice is shown to be sufficient. 
 Numerical simulations indicate that extending to three or more parameter choices is never beneficial.
	\end{abstract}
	\section{Introduction}
	In many Internet of things (IoT) and sensor networks, the sensors may not communicate directly with the decision center due to limited resources or environmental effects. This motivates us to consider  multi-hop networks where the sensor can communicate to the decision center  only via a relay.  In certain scenarios, the relays also wish to decide on the hypothesis, for example to faster raise alarms. In such distributed hypothesis testing problems, the relays and the receiver have to decide on a binary hypothesis to determine the joint distributions underlying all terminals' observations including their own. In particular, maximizing the accuracy of any taken decision under imposed \emph{communication rate constraints} is an important concern in many applications related to security, health monitoring, or incident-detection. In these applications, often the error under the alternative hypothesis corresponding to a \emph{missed detection} is more critical than the error under the null hypothesis corresponding to \emph{false alarms}. We thus  aim at maximizing the exponential decays of the missed detection probabilities under given thresholds on the  false alarm probabilities. As we shall see, a particular challenge arises when the relay and the decision center have different thresholds on the tolerable false-alarm probabilities.

Most information-theoretic works on distributed hypothesis testing focus on \emph{maximum rate constraints}\cite{Ahlswede,Han,Amari,Wagner,Michele2,zhao2018distributed}. \emph{Expected rate constraints} were introduced in \cite{MicheleVLconf,JSAIT}, which also characterized the maximum  error exponents  for single-sensor single-decision center setups in the special case of testing-against independence. The optimal coding and decision scheme in \cite{MicheleVLconf,JSAIT} chooses an  event  $\mathcal{S}_n$ of probability close  to the permissible type-I error probability $\epsilon$. Under this event, the transmitter   sends  a single flag bit to the decision center, which then decides on the hypothesis $\mathcal{H}=1$. Otherwise, the transmitter and the receiver run the optimal scheme under the maximum rate constraints \cite{Ahlswede,Han}. The described scheme achieves same type-II error exponent  as in \cite{Ahlswede,Han}, but with a  communication rate reduced by  the factor of  $(1-\epsilon)$. 
Similar conclusions also hold for more complicated  networks with multiple communication links, as we showed   in \cite{HWS20} at hand of the partially-cooperating multi-access network with two sensors. 

In this paper, we consider the two-hop network, where the observations at the transmitter $X^n$, the relay $Y^n$, and the receiver $Z^n$ form a Markov chain $X^n\to Y^n \to Z^n$.  Such a Markov chain often occurs simply because the transmitter is closer to the relay than to the receiver. Under maximum rate-constraints, the optimal exponents at the relay and the receiver were characterized in \cite{Michele,Vincent}. We show that when both the transmitter and the relay have same $\epsilon_1=\epsilon_2$, then  under  expected rate constraints one can boost both rates  by a factor $(1-\epsilon)^{-1}$ as compared to a maximum rate-constraint. 
The case $\epsilon_1 \neq \epsilon_2$ differs in various ways. Firstly, our set of  achievable exponent pairs indicates a tradeoff between the relay's and the receiver's exponents. Secondly, a more complicated coding and decision scheme is required. Specifically, we propose a strategy   
where the transmitter chooses three events, and depending on the event, applies either a  degenerate single-flagbit strategy or  the scheme in \cite{Michele} with one of two different choices of parameters and rates, depending on the transmitter's observation.  Extending to more than three events (i.e., to more than two parameter and rate choices for the scheme in \cite{Michele}) however does not seem to yield further improvements.

\textit{Notation:}
	We follow the notation in \cite{ElGamal},\cite{JSAIT}. In particular, we use sans serif font for bit-strings: e.g., $\m$ for a deterministic and $\M$ for a random bit-string. We let  $\mathrm{string}(m)$ denote the shortest bit-string representation of a positive integer  $m$, and for any bit-string $\m$ we let  $\mathrm{len}(\m)$  and $\mathrm{dec}(\m)$ denote its length and its corresponding positive integer. In addition, $\mathcal{T}_{\mu}^{(n)}$ denotes the strongly typical set given by \cite[Definition 2.8]{Csiszarbook}. 
	
	\section{System Model}

\begin{figure}[htbp]
		\centerline{\includegraphics[width=7.5cm, scale=0.5]{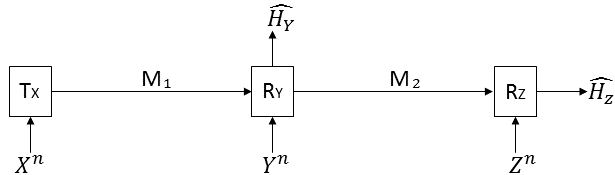}}
		\caption{Cascaded two-hop setup with two decision centers.}
		\label{fig:Cascaded}
	\end{figure}

	Consider the distributed hypothesis testing problem in Fig.~\ref{fig:Cascaded} under the Markov chain  
		\begin{equation}\label{eq:Mc}
		X^n \to Y^n \to Z^n
	\end{equation}
and  in the special case of testing against independence, i.e., depending on the binary hypothesis $\mathcal{H}\in\{0,1\}$, the tuple $(X^n,Y^n,Z^n)$ is distributed as:
	\begin{subequations}\label{eq:dist}
	\begin{IEEEeqnarray}{rCl}
		& &\textnormal{under } \mathcal{H} = 0: (X^n,Y^n,Z^n) \sim \textnormal{i.i.d.} \, P_{XY}\cdot P_{Z|Y} ; \label{eq:H0_dist}\IEEEeqnarraynumspace\\
		& &\textnormal{under } \mathcal{H} = 1: (X^n,Y^n,Z^n) \sim \textnormal{i.i.d.} \, P_{X}\cdot P_{Y}\cdot P_{Z}
	\end{IEEEeqnarray} 
	\end{subequations}
	for given probability mass functions (pmfs) $P_{XY}$ and $P_{Z|Y}$.
	
The system consists of a transmitter T$_{X}$, a relay R$_{Y}$, and a receiver R$_Z$. The transmitter T$_{X}$ observes the source sequence $X^n$ and sends its bit-string message $\M_1 = \phi_1^{(n)}(X^n)$ to R$_Y$, where the encoding function is of the form $\phi_1^{(n)} : \mathcal{X}^n \to \{0,1\}^{\star}$ and satisfies the \emph{expected} rate constraint
	\begin{equation}\label{eq:Rate1}
		\mathbb{E}\left[\mathrm{len}\left(\M_1\right)\right]\leq nR_1.
	\end{equation} 
	The relay R$_Y$ observes the source sequence $Y^n$ and with the message $\M_1$ received from T$_{X}$, it produces a guess  $\hat{\mathcal{H}}_Y$ of the hypothesis ${\mathcal{H}}$ using a decision function $g_1^{(n)} : \mathcal{Y}^n \times \{0,1\}^{\star} \to \{0,1\}$:
	\begin{equation}
		\hat{\mathcal{H}}_Y = g_1^{(n)}\left(\M_1,Y^n\right) \;   \in\{0,1\}.
	\end{equation}
	Relay R$_Y$ also computes a bit-string message $\M_2 = \phi_2^{(n)}\left(Y^n,\M_1\right)$ using some encoding function $\phi_2^{(n)}: \mathcal{Y}^n\times\{0,1\}^{\star}\to\{0,1\}^{\star}$ that satisfies the expected rate constraint
	\begin{equation}\label{eq:Rate2}
		\mathbb{E}\left[\mathrm{len}\left(\M_2\right)\right]\leq nR_2.
	\end{equation} Then it sends $\M_2$ to the receiver R$_Z$, which guesses hypothesis $\mathcal{H}$ using   its  observation $Z^n$ and the received message $\M_2$, i.e.,  using a decision function $g_2^{(n)} : \mathcal{Z}^n \times \{0,1\}^{\star} \to \{0,1\}$, it produces the guess:
	\begin{equation}
		\hat{\mathcal{H}}_Z = g_2^{(n)}\left(\M_2,Z^n\right) \;   \in\{0,1\}.
	\end{equation}
	
	The  goal is to design encoding and decision functions such that their type-I error probabilities 
	\begin{IEEEeqnarray}{rCl}
		\alpha_{1,n} &\triangleq& \Pr[\hat{\mathcal{H}}_Y = 1|\mathcal{H}=0]\\
		\alpha_{2,n} &\triangleq& \Pr[\hat{\mathcal{H}}_Z = 1|\mathcal{H}=0]
	\end{IEEEeqnarray}
	stay below given thresholds $\epsilon_1 > 0$, $\epsilon_2 > 0$ and the type-II error probabilities
	\begin{IEEEeqnarray}{rCl}
		\beta_{1,n} &\triangleq& \Pr[\hat{\mathcal{H}}_Y = 0|\mathcal{H}=1]\\
		\beta_{2,n} &\triangleq& \Pr[\hat{\mathcal{H}}_Z = 0|\mathcal{H}=1]
	\end{IEEEeqnarray}
	decay to 0 with largest possible exponential decay.  
		
	\begin{definition} Fix maximum type-I error probabilities $\epsilon_1,\epsilon_2 \in [0,1]$ and rates $R_1,R_2 \geq 0$. The exponent pair $(\theta_1,\theta_2)$ is called \emph{$(\epsilon_1,\epsilon_2)$-achievable} if there exists a sequence of encoding and decision functions $\{\phi_1^{(n)},\phi_2^{(n)},g_1^{(n)},g_2^{(n)}\}_{n\geq 1}$ satisfying $\forall i \in \{1,2\}$:
		\begin{IEEEeqnarray}{rCl}
			\mathbb{E}[\text{len}(\M_i)] &\leq& nR_i, \\
			\varlimsup_{n \to \infty}\alpha_{i,n} & \leq& \epsilon_i,\label{type1constraint1}\\ 
			\label{thetaconstraint}
			\varliminf_{n \to \infty}  {1 \over n} \log{1 \over \beta_{i,n}} &\geq& \theta_i.
		\end{IEEEeqnarray}
	\end{definition}
\begin{definition}
The closure of the set of all $(\epsilon_1,\epsilon_2)$-achievable exponent pairs $(\theta_{1},\theta_{2})$ is called the \emph{$(\epsilon_1,\epsilon_2)$-exponents region} and is denoted by $\mathcal{E}^*(R_1,R_2,\epsilon_1,\epsilon_2)$. 

The maximum  exponents that are achievable at  each  of the two decision centers  are also of interest:
\begin{IEEEeqnarray}{rCl}
	\theta^*_{1,\epsilon_1}(R_1) &:=& \max \{\theta_{1} \colon \, (\theta_{1},\theta_{2}) \in \mathcal{E}^*(R_1,R_2,\epsilon_1,\epsilon_2)  \nonumber \\
	&& \hspace{1.9cm} \textnormal{ for some } \epsilon_2>0,\theta_2 \geq 0\} \label{eq:maxtheta1}\\
	\theta^*_{2,\epsilon_2}(R_1,R_2) &:=& \max\{\theta_{2} \colon \, (\theta_{1},\theta_{2}) \in \mathcal{E}^*(R_1,R_2,\epsilon_1,\epsilon_2) \nonumber \\
	&& \hspace{1.9cm} \textnormal{ for some } \epsilon_1>0, \theta_1 \geq 0\}.\IEEEeqnarraynumspace\label{eq:maxtheta2}
\end{IEEEeqnarray}
\end{definition}

		\begin{remark}
The  multi-hop hypothesis testing setup of Fig.~\ref{fig:Cascaded} and Equations~\eqref{eq:dist} was also considered in   \cite{Michele} and \cite{Vincent}, but under   \emph{maximum rate constraints}:
\begin{equation}\label{eq:FixRates}
\len(\M_i) \leq nR_i,	\qquad i\in\{1,2\},
\end{equation} 
instead of the \emph{expected rate constraints} \eqref{eq:Rate1} and \eqref{eq:Rate2}.

As shown in \cite{Vincent}, for any rates $R_1,R_2\geq 0$ and permissible type-I error probabilities $\epsilon_1,\epsilon_2 \in[0,1/2]$, the  exponents region 	under the maximum-rate constraints \eqref{eq:FixRates} is:
	\begin{IEEEeqnarray}{rCl}
		\mathcal{E}_{\fix}^*(R_1,R_2,\epsilon_1,\epsilon_2) = \{(\theta_{1},\theta_{2}) : \theta_1 &\leq& \theta_{1,\epsilon_1,\fix}^{*}\left(R_1\right),\\ \theta_{2} &\leq& \theta_{2,\epsilon_2,\fix}^{*}\left(R_1,R_2\right)\},\IEEEeqnarraynumspace
	\end{IEEEeqnarray}
	where
	\begin{IEEEeqnarray}{rCl}
		\theta_{1,\epsilon_1,\fix}^{*}\left(R_1\right) &=& \max\limits_{\substack{P_{U_1|X}\colon \\R_1 \geq I\left(U_1;X\right)}} I\left(U_1;Y\right)\\
		\theta_{2,\epsilon_2,\fix}^{*}\left(R_1,R_2\right) &=& \theta_{1,\epsilon_1,\fix}^{*}\left(R_1\right) + \max\limits_{\substack{P_{U_2|Y}\colon \\R_2 \geq I\left(U_2;Y\right)}}  I\left(U_2;Z\right) \IEEEeqnarraynumspace
	\end{IEEEeqnarray}
	and the mutual information quantities are calculated using the joint pmfs $P_{U_1XY} \triangleq P_{U_1|X}P_{XY}$ and $P_{U_2YZ} \triangleq P_{U_2|Y}P_{YZ}$.

In the following subsection~\ref{sec:scheme_basic} we  present a coding and decision scheme  that achieves $\mathcal{E}_{\fix}^*(R_1,R_2,\epsilon_1,\epsilon_2)$. It is a simplification of the scheme in \cite{Michele}.

	\end{remark}
	
	\section{Coding and Decision Schemes}\label{sec:schemes}

In Subsection~\ref{sec:scheme_basic}, we  present a basic two-hop hypothesis testing scheme, which we obtain by simplifying the general scheme in \cite{Michele} and which suffices to achieve the exponents region $\mathcal{E}_{\fix}^*$ under maximum rate constraints. 

For the setup with  \emph{expected} rate constraints studied in this paper, in Subsections~\ref{sec:scheme_same}--\ref{sec:scheme1_larger} we propose to use different versions of this two-hop scheme (with different parameters and different communication rates) depending on the transmitter's observation $x^n$, where for certain sequences $x^n$ we even apply degenerate versions of the scheme where only zero-rate flag-bits  are sent over one or both communication links.  Notice that in principle, we could apply a different set of parameters for each observation  $x^n \in \mathcal{X}^n$. Our numerical examples however indicate that without loss in optimality one can restrict to only  one or two  parameter choices and an additional degenerate version of the scheme with zero communication rates on both links. As proved by the scheme in Subsection~\ref{sec:scheme_same} and  Theorem~\ref{thm1} a single parameter choice suffices when $\epsilon_1=\epsilon_2$. For  $\epsilon_1\neq \epsilon_2$ two parameter choices are strictly better as we show in our numerical simulations in Section~\ref{sec:numerical}. More choices seem unnecessary.

\subsection{A basic two-hop coding and decision scheme \cite{Michele}}\label{sec:scheme_basic}
We revisit a simplified version of  the scheme  in \cite{Michele}, which achieves the exponents region under maximum rate constraints $\mathcal{E}_{\fix}^*(R_1,R_2,\epsilon_1,\epsilon_2)$ for any $\epsilon_1,\epsilon_2$. 

Fix a blocklength $n$ and choose the following parameters: a small positive number $\mu >0$, conditional pmfs $P_{U_1|X}$ and $P_{U_2|Y}$. In the following, all mutual informations will be evaluated according to the joint pmf $P_{XYZU_1U_2}:=P_{X}P_{Y|X}P_{Z|Y}P_{U_1|X}P_{U_2|Y}$. 

Randomly generate the codebooks
\begin{IEEEeqnarray}{rCl}
	\mathcal C_{U_1} &\triangleq& \left\{u_1^{n}(m_1): m_1 \in \left\{1,\cdots,2^{n\left(I\left(U_1;X\right) + \mu \right)}\right\}\right\}\\
	\mathcal C_{U_2} &\triangleq& \left\{u_2^{n}(m_2): m_2 \in  \left\{ 1, \cdots,2^{n\left(I\left(U_2;Y\right) + \mu \right)}\right\}\right\},
\end{IEEEeqnarray}
by drawing all entries i.i.d. according to the marginal pmfs $P_{U_1}$ and $P_{U_2}$.\\

\underline{{\Tx:}} Assume it observes $X^n=x^n$.  If  $x^n \in \mathcal{T}_{\mu}^{(n)} (P_X)$, it looks for indices $m_1$ satisfying $\left(u_1^{n}(m_1),x^n\right) \in \mathcal{T}_{\mu}^{(n)}(P_{U_1X})$, randomly picks one of these indices, and sends its corresponding bit-string 
\begin{equation}
\M_1 = [\mathrm{string} (m_1)].
\end{equation}
If no such index exists or if $x^n \notin \mathcal{T}_{\mu}^{(n)} (P_X)$,
then \Tx \,sends
\begin{equation}
\M_1=[0].
\end{equation}

\underline{{\Rel:}} Assume it observes  $Y^n = y^n$ and receives the bit-string message $\M_1=\m_1$.

If $\m_1 = [0]$,  then  \begin{equation}
\hat{\mathcal{H}}_Y = 1 \qquad \textnormal{and} \qquad \M_2 = [0].\end{equation}

Else  it checks if  $\left(u_1^{n}(\m_1),y^n\right)\in\mathcal{T}_{\mu}^{(n)}(P_{U_1Y})$. If the check is successful {\Rel} declares   $\hat{\mathcal{H}}_Y = 0$; otherwise it declares   $\hat{\mathcal{H}}_Y = 1$ and sends $\M_2=[0]$.\vspace{1mm}

 If $\hat{\mathcal{H}}_Y=0$, {\Rel} next looks for  indices $m_2$ satisfying $\left(u_2^{n}(m_2),y^n\right)\in\mathcal{T}_{\mu}^{(n)}(P_{U_2Y})$, randomly picks one of them and sends 
 \begin{equation}
\M_2 = \mathrm{string} (m_2)
\end{equation}
 to the receiver. 
 
 If no such index $m_2$  exists, {\Rel}  directly sends  string \begin{equation}
 \M_2=[0].
 \end{equation}
 
\underline{{\Rec:}} Assume it observes the sequence $Z^n = z^n$ and receives message $\M_2=\m_2$. 

If $\m_2 = [0]$, it declares $\hat{\mathcal{H}}_Z=1$. \vspace{1mm}

Else it sets $m_2=\textnormal{dec}(\m_2)$, and checks if $\left(u_2^{n}(m_2),z^n\right)\in\mathcal{T}_{\mu}^{(n)}(P_{U_2Z})$. It declares $\hat{\mathcal{H}}_Z = 0$ if the check succeeds, and  $\hat{\mathcal{H}}_Z = 1$ otherwise.

 In the following subsections, we explain how to employ this basic scheme in a variable-length coding framework. 
\subsection{Variable-length coding for  $\epsilon_1=\epsilon_2$}\label{sec:scheme_same}

We employ  only a single version of the two-hop scheme, and combine it with  a degenerate scheme that has zero communication rates over both links. Specifically, as for the point-to-point setup in \cite{JSAIT}, we choose a subset $\mathcal{S}_n \subseteq \mathcal{T}_\mu^{(n)}(P_X)$ of probability 
\begin{IEEEeqnarray}{rCl}
	\mathrm{Pr}\left[X^n \in \mathcal{S}_n\right] &=& \epsilon_2 -\mu = \epsilon_1-\mu,
\end{IEEEeqnarray}
for some small number $\mu>0$. 

Whenever $X^n \in \mathcal{S}_n$, \Tx \,and \Rel \,both send
\begin{equation}\label{eq:zerorate}
\M_1=\M_2=[0]
\end{equation}
and \Rel \,and \Rec \,decide on 
\begin{equation}\label{eq:HyHz1}
\hat{\mathcal{H}}_Y=\hat{\mathcal{H}}_Z=1.
\end{equation}

Whenever $X^n \notin \mathcal{S}_n$,
the terminals {\Tx}, {\Rel}, {\Rec} all follow the basic two-hop scheme  in Subsection \ref{sec:scheme_basic} for parameters $\mu, P_{U_1|X}, P_{U_2|Y}$ satisfying 
\begin{IEEEeqnarray}{rCl}
	R_1 &=& \left( 1- \epsilon_1 + \mu \right)\left(I(U_1;X) + 2\mu\right)\\
	R_2 &=& \left( 1-\epsilon_2 + \mu \right)\left(I(U_2;Y) + 2\mu\right).
	\end{IEEEeqnarray}
The factors $(1-\epsilon_1+\mu)$ and $(1-\epsilon_2+\mu )$ in front of the mutual information terms represent the gain obtained by \emph{expected} rate constraints, because with probability $\epsilon_1-\mu=\epsilon_2-\mu$   in our scheme both messages $\M_1$ and $\M_2$ are of zero rate, see \eqref{eq:zerorate}.

In Appendix~\ref{app1}, we prove that the presented scheme achieves the error exponents in Eq.~\eqref{eq:E1} of Theorem~\ref{thm1} when $n\to \infty$ and $\mu \downarrow 0$.

\subsection{Variable-length coding for $\epsilon_2 > \epsilon_1$}\label{sec:scheme_larger}

We employ two versions of the basic two-hop scheme as we will explain shortly. Moreover, we again  choose a subset $\mathcal{S}_n \subseteq \mathcal{T}_\mu^{(n)}(P_X)$ of probability 
\begin{IEEEeqnarray}{rCl}
	\mathrm{Pr}\left[X^n \in \mathcal{S}_n\right] &=& \epsilon_1 -\mu ,
\end{IEEEeqnarray}
and all terminals {\Tx}, {\Rel}, {\Rec} apply the degenerate scheme in \eqref{eq:zerorate}--\eqref{eq:HyHz1} whenever 
 $X^n \in \mathcal{S}_n$. 

We now partition the remaining set $\mathcal{X}^n \backslash \mathcal{S}_n$ into two disjoint sets $\mathcal{D}_n'$ and $\mathcal{D}_n''$
\begin{equation}
\mathcal{D}_n' \cup \mathcal{D}_n''=\mathcal{X}^n \backslash \mathcal{S}_n \quad \textnormal{and} \quad \mathcal{D}_n' \cap \mathcal{D}_n''=\emptyset
\end{equation}
such that 
\begin{IEEEeqnarray}{rCl}
	\mathrm{Pr}\left[X^n \in \mathcal{D}_n'\right] &=& 1- \epsilon_2+\mu\\
	\mathrm{Pr}\left[X^n \in \mathcal{D}_n''\right] &=& \epsilon_2-\epsilon_1.
\end{IEEEeqnarray}
We further split $R_1=R_1'+R_1''$  for $R_1',R_1''>0$.

Then, whenever $x^n \in \mathcal{D}_n'$, all terminals
 {\Tx, \Rel, \Rec} follow the basic two-hop scheme  for a set of parameters $\mu, P_{U_1'|X}, P_{U_2'|Y}$ satisfying 
\begin{IEEEeqnarray}{rCl}
	R_1' &=&(1-\epsilon_2 +\mu) (I(U_1';X) + 2\mu)\\
	R_2 &=& (1-\epsilon_2 +\mu)(I(U_2';Y) + 2\mu). 
\end{IEEEeqnarray}
To inform the relay and the receiver about the event $x^n \in \mathcal{D}_n'$, both {\Tx} and {\Rel} add $[1,0]$-flag bits at the beginning of their communication to {\Rel} and {\Rec}, respectively.  (Notice that two additional   bits do not change  the rate of communication.)

For  $x^n \in \mathcal{D}_n''$,
the transmitter and the  relay still follow the basic two-hop scheme  in Subsection~\ref{sec:scheme_basic} but now  for a different parameter choice  $\mu, P_{U_1''|X}$ satisfying 
\begin{IEEEeqnarray}{rCl}
	R_1'' &=&(\epsilon_2 -\epsilon_1)(I(U_1'';X) + 2\mu),
\end{IEEEeqnarray}
and where {\Tx}  additionally sends the $[1,1]$-flag as part of $\M_1$ to \Rel, which simply relays this flag $\M_2=[1,1]$ without adding additional information. 
Upon observing $\M_2=[1,1]$, {\Rec} immediately  declares $\hat{\mathcal{H}}_Z=1$. 

In Appendix~\ref{app2}, we prove that the presented scheme achieves the error exponents in Eq.~\eqref{eq:E2} of Theorem~\ref{thm1} when $n\to \infty$ and $\mu \downarrow 0$.

\subsection{Variable-length coding for $\epsilon_1 > \epsilon_2$}\label{sec:scheme1_larger}

In this case, we  employ two full versions of the basic two-hop scheme. Moreover, we again  choose a subset $\mathcal{S}_n \subseteq \mathcal{T}_\mu^{(n)}(P_X)$ of probability
\begin{IEEEeqnarray}{rCl}
	\mathrm{Pr}\left[X^n \in \mathcal{S}_n\right] &=& \epsilon_2 -\mu,
\end{IEEEeqnarray}
and partition the remaining subset of $\mathcal{X}^n$ into two disjoint sets $\mathcal{D}_n'$ and $\mathcal{D}_n''$
\begin{equation}
\mathcal{D}_n' \cup \mathcal{D}_n''=\mathcal{X}^n \backslash \mathcal{S}_n \quad \textnormal{and} \quad \mathcal{D}_n' \cap \mathcal{D}_n''=\emptyset
\end{equation}
such that 
\begin{IEEEeqnarray}{rCl}
\mathrm{Pr}\left[X^n \in \mathcal{D}_n'\right] &=& 1- \epsilon_1+\mu\\
\mathrm{Pr}\left[X^n \in \mathcal{D}_n''\right] &=& \epsilon_1-\epsilon_2.
\end{IEEEeqnarray}
We further split $R_1=R_1'+R_1''$ and $R_2=R_2'+R_2''$ for $R_1',R_1'',R_2',R_2''>0$.

Whenever $X^n \in \mathcal{S}_n$,  \Tx, \Rel, and {\Rec}, all apply the degenerate scheme in \eqref{eq:zerorate}--\eqref{eq:HyHz1}.

Whenever $X^n \in \mathcal{D}_n'$, all terminals
\Tx, \Rel, and {\Rec} follow the basic two-hop scheme for a choice of parameters $\mu, P_{U_1'|X}, P_{U_2'|Y}$ satisfying 
\begin{IEEEeqnarray}{rCl}
	R_1' &=&(1-\epsilon_1+\mu)(I(U_1';X) + 2\mu)\\
	R_2' &=& (1-\epsilon_1+\mu)I(U_2';Y) + 2\mu).
\end{IEEEeqnarray}
Additionally, \Tx \,and \Rel \, add $[1,0]$-flag bits to their messages $\M_1$ and $\M_2$  to indicate to  \Rel \,and \Rec \,that $X^n \in \mathcal{D}_n'$.

Whenever $X^n \in \mathcal{D}_n''$,
all terminals \Tx, \Rel, and {\Rec} mostly follow the basic two-hop scheme but now  for parameters $\mu, P_{U_1''|X}, P_{U_2''|Y}$ satisfying 
\begin{IEEEeqnarray}{rCl}
	R_1'' &=&(\epsilon_1-\epsilon_2)(I(U_1'';X) + 2\mu)\\
		R_2'' &=&(\epsilon_1-\epsilon_2)( I(U_2'';Y) + 2\mu).
\end{IEEEeqnarray}
The only exceptions are that  \Tx \,and \Rel \,add a $[1,1]$-flag to their messages  $\M_1$ and $\M_2$ to indicate to \Rel \,and \Rec \, that $X^n \in \mathcal{D}_n''$, and  that {\Rel} \emph{always} declares $\hat{\mathcal{H}}_Y = 1$   upon observing this $[1,1]$-flag in $\M_1$, irrespective of the remaining bits of $\M_1$ or its observation $Y^n$. Besides this decision, {\Rel} however follows the protocol of the basic two-hop scheme which forces it to compute a tentative decision  $\hat{\mathcal{H}}_Y ''$, which determines its communication to {\Rec}. (In particular, if $\hat{\mathcal{H}}_Y''=1$, {\Rel} sends only the $[1,1]$-flag to {\Rec} so that {\Rec} immediately declares $\hat{\mathcal{H}}_Z=1$.) Notice that while {\Rel} can ignore the tentative decision $\hat{\mathcal{H}}_Y''$ because of its larger permissible type-I error probability $\epsilon_1>\epsilon_2$, this decision is important for {\Rec} so that this latter can satisfy  its constraint on the type-I probability $\epsilon_2$.

In a similar way to the previous schemes, it can be shown that this  scheme achieves the error exponents in  Eq.~\eqref{eq:E3} of Theorem~\ref{thm1} when $n\to \infty$ and $\mu \downarrow 0$. Details are presented in Appendix~\ref{app3}.

	\section{Results on the Exponents Region}
		
Our main result provides   inner bounds to the exponent region $\mathcal{E}^*(R_1,R_2,\epsilon_1,\epsilon_2)$ achieved by the schemes presented in the preceding Section~\ref{sec:schemes}. The theorem further provides an exact characterization of exponents region $\mathcal{E}^*(R_1,R_2,\epsilon_1,\epsilon_2)$ when $\epsilon_1=\epsilon_2$.

	\begin{thm}\label{thm1}
	If $\epsilon_1=\epsilon_2$,   the $(\epsilon_1,\epsilon_2)$-exponents region $\mathcal{E}^*(R_1,R_2,\epsilon_1,\epsilon_2)$ \emph{is the set of} all ($\theta_{1},\theta_{2}$) pairs  satisfying
\begin{subequations}\label{eq:E1}
	\begin{IEEEeqnarray}{rCl}
		\theta_{1} &\leq&I(U_1;Y),\\
		\theta_{2} &\leq& I(U_1;Y) + I(U_2;Z),
	\end{IEEEeqnarray}
	for some conditional pmfs $P_{U_1|X},P_{U_2|Y}$ so that
\begin{IEEEeqnarray}{rCl}
R_1 &\geq& (1-\epsilon_1)I(U_1;X), \\
R_2 &\geq&  (1- \epsilon_2)I(U_2;Y),
\end{IEEEeqnarray}
\end{subequations}
and where the mutual information quantities are calculated using the joint pmfs $P_{U_1XY} \triangleq P_{U_1|X}P_{XY}$ and $P_{U_2YZ} \triangleq P_{U_2|Y}P_{YZ}$.

		If $\epsilon_1<\epsilon_2$, the $(\epsilon_1,\epsilon_2)$-exponents region $\mathcal{E}^*(R_1,R_2,\epsilon_1,\epsilon_2)$ \emph{contains} all ($\theta_{1},\theta_{2}$) pairs that satisfy 
\begin{subequations}\label{eq:E2}
	\begin{IEEEeqnarray}{rCl}
		\theta_{1} &\leq& \min\{I(U_1';Y),I(U_1'';Y)\}, \\
		\theta_{2} &\leq& I(U_1';Y) + I(U_2';Z),
	\end{IEEEeqnarray}
	for some conditional pmfs $P_{U_1'|X},P_{U_1''|X},P_{U_2'|Y}$ so that
\begin{IEEEeqnarray}{rCl}
R_1 &\geq& (1-\epsilon_2)I(U_1';X) + (\epsilon_{2} - \epsilon_1)I(U_1'';X), \\
R_2 &\geq&  (1- \epsilon_2)I(U_2';Y),
\end{IEEEeqnarray}
\end{subequations}
and where the mutual information quantities are calculated using the joint pmfs $P_{U_1'XY} \triangleq P_{U_1'|X}P_{XY}$, $P_{U_1''XY} \triangleq P_{U_1''|X}P_{XY}$, and $P_{U_2'YZ} \triangleq P_{U_2'|Y}P_{YZ}$.

		If $\epsilon_1>\epsilon_2$,   the $(\epsilon_1,\epsilon_2)$-exponents region $\mathcal{E}^*(R_1,R_2,\epsilon_1,\epsilon_2)$ \emph{contains} all ($\theta_{1},\theta_{2}$) pairs that satisfy 
\begin{subequations}\label{eq:E3}
\begin{IEEEeqnarray}{rCl}
 \theta_{1} &\leq& I(U_1';Y), \\
  \theta_{2} &\leq& \min\{I(U_1';Y)+ I(U_2';Z),I(U_1'';Y)+ I(U_2'';Z)\}, \IEEEeqnarraynumspace
		\end{IEEEeqnarray}
	for some conditional pmfs $P_{U_1'|X},P_{U_1''|X},P_{U_2'|Y},P_{U_2''|Y}$ so that 
\begin{IEEEeqnarray}{rCl}
R_1 &\geq& (1-\epsilon_1)I(U_1';X) + (\epsilon_1 - \epsilon_2)I(U_1'';X), \\
R_2 &\geq&  (1-\epsilon_1)I(U_2';Y) + (\epsilon_{1}- \epsilon_2)I(U_2'';Y),
\end{IEEEeqnarray}%
\end{subequations}
and where the mutual information quantities are calculated using the joint pmfs $P_{U_1'XY} \triangleq P_{U_1'|X}P_{XY}$, $P_{U_1''XY} \triangleq P_{U_1''|X}P_{XY}$, $P_{U_2'YZ} \triangleq P_{U_2'|Y}P_{YZ}$, and $P_{U_2''YZ} \triangleq P_{U_2''|Y}P_{YZ}$.
	\end{thm}
	\begin{IEEEproof} Achievability results  are based on the schemes in Section~\ref{sec:schemes}, see Appendices \ref{app1}, \ref{app2}, and \ref{app3} for the analyses. For $\epsilon_{1}=\epsilon_{2}$ the converse is proved  in Appendix~\ref{converse}.
	\end{IEEEproof}
	
\subsection{Numerical Simulations}\label{sec:numerical}

In this section, we illustrate the benefits of variable-length coding as opposed to fixed-length coding (or the benefits of having the relaxed expected rate constraints in \eqref{eq:Rate1} and \eqref{eq:Rate2} instead of the more stringent maximum rate-constraints \eqref{eq:FixRates}). We also show for $\epsilon_2 \neq \epsilon_1$ the benefits of having two auxiliary random variables $U_1'$ and $U_1''$ in \eqref{eq:E2}--\eqref{eq:E3} instead of only a single random variable, which is equivalent to applying the basic two-hop scheme for two  parameter choices (depending on $X^n$) and not just one.  And finally, for $\epsilon_2 < \epsilon_1$, we illustrate the benefits of having both $U_2'$ and $U_2''$ in \eqref{eq:E3}, which stems from applying two full versions of the basic two-hop scheme in Subsection~\ref{sec:scheme_basic}.

Throughout this section we consider the following example. 
	Let $X,S,T$ be  independent Bernoulli random variables of parameters $p_{X}=0.4,p_S=0.8,p_T=0.8$ and set $Y=X \oplus T$ and $Z=Y \oplus S$. 
	
\begin{figure}[htbp]
			\begin{tikzpicture} [every pin/.style={fill=white},scale=.875]
				\begin{axis}[scale=1,
					width=.93\columnwidth,
					scale only axis,
					xmin=0.3,
					xmax=0.8,
					xmajorgrids,
					xlabel={$R_1=R_2=R$},
					ymin=0.1,
					ymax=0.5,
					ymajorgrids,
					ylabel={$\theta$},
					axis x line*=bottom,
					axis y line*=left,
					legend pos=north west,
					legend style={draw=none,fill=none,legend cell align=left, font=\normalsize}
					]

					\addplot[color=blue,solid,line width=2pt]
					table[row sep=crcr]{		
						0.25		0.180209513851953\\
						0.275		0.197246942523415\\
						0.3			0.214085790120879\\
						0.325		0.230720318421095\\
						0.35		0.247144481768313\\
						0.375		0.263351896994993\\
						0.4			0.279335808627650\\
						0.425		0.295089048353326\\
						0.45		0.310603987436604\\
						0.475		0.325872480392762\\
						0.5			0.340885797698503\\
						0.525		0.355634544595247\\
						0.55		0.370108562016254\\
						0.575		0.384296804198065\\
						0.6			0.398187185378442\\
						0.625		0.411766384738483\\
						0.65		0.425019593733904\\
						0.675		0.437930181968939\\
						0.7			0.450479244552176\\
						0.725		0.462644971066085\\
						0.75		0.474401734844207\\
						0.775		0.485718721104490\\
						0.8			0.496557744856998\\
						0.825		0.506869521480579\\
						0.85		0.516586621708389\\
						0.875		0.525607983563066\\
						0.9			0.534175556817941\\
						0.925		0.540234297422347\\
						0.95		0.541988613910995\\
						0.975		0.541988627540627\\
						1			0.541988620724640\\};
					\addlegendentry{$\theta_{2,\epsilon}^*$}

					\addplot[color=blue,dashdotted,line width=2pt]
					table[row sep=crcr]{
						0.25		0.171618047255172\\
						0.275		0.187900576023716\\
						0.3			0.204006626066673\\
						0.325		0.219931419036764\\
						0.35		0.235669939363905\\
						0.375		0.251216912944721\\
						0.4			0.266566782771540\\
						0.425		0.281713680892380\\
						0.45		0.296651395938766\\
						0.475		0.311373335255028\\
						0.5			0.325872480392762\\
						0.525		0.340141334370662\\
						0.55		0.354171858603691\\
						0.575		0.367955396717514\\
						0.6			0.381482581493401\\
						0.625		0.394743219991748\\
						0.65		0.407726148261809\\
						0.675		0.420419049522989\\
						0.7			0.432808213798348\\
						0.725		0.444878223264951\\
						0.75		0.456611523819971\\
						0.775		0.467987827062266\\
						0.8			0.478983245460232\\
						0.825		0.489568986341129\\
						0.85		0.499709267826483\\
						0.875		0.509357741282936\\
						0.9			0.518450689681937\\
						0.925		0.526891915394668\\
						0.95		0.534508215452524\\
						0.975		0.540400161997789\\
						1			0.541988627549646\\};
					\addlegendentry{$\theta_{2,\epsilon,\fix}^*$}
					\addplot[color=teal,solid,line width=2pt]
					table[row sep=crcr]{
						0.25		0.0898036524235792\\
						0.275		0.0982877008342298\\
						0.3			0.106671468550581\\
						0.325		0.114952034477960\\
						0.35		0.123126318852111\\
						0.375		0.131191067443298\\
						0.4			0.139142833242811\\
						0.425		0.146977955076165\\
						0.45		0.154692532429851\\
						0.475		0.162282395565877\\
						0.5			0.169743069706874\\
						0.525		0.177069731668621\\
						0.55		0.184257156741615\\
						0.575		0.191299652791754\\
						0.6			0.198190977321526\\
						0.625		0.204924231371814\\
						0.65		0.211491721243744\\
						0.675		0.217884774349693\\
						0.7			0.224093487692167\\
						0.725		0.230106373810154\\
						0.75		0.235909843829443\\
						0.775		0.241487417598940\\
						0.8			0.246818444674268\\
						0.825		0.251875867010773\\
						0.85		0.256621857157859\\
						0.875		0.260997778363860\\
						0.9			0.265314244122168\\
						0.925		0.267659426301970\\
						0.95		0.267659411992428\\
						0.975		0.267659426246069\\
						1			0.267659425886040\\};
					\addlegendentry{$\theta_{1,\epsilon}^*$}
					
					\addplot[color=teal,dashed,line width=2pt]
					table[row sep=crcr]{
						0.25		0.0855248958260029\\
						0.275		0.0936337032344023\\
						0.3			0.101653401799731\\
						0.325		0.109581559976988\\
						0.35		0.117415623895128\\
						0.375		0.125152906181165\\
						0.4			0.132790573152924\\
						0.425		0.140325630051096\\
						0.45		0.147754903896272\\
						0.475		0.155075023444565\\
						0.5			0.162282395565877\\
						0.525		0.169373177166683\\
						0.55		0.176343241501810\\
						0.575		0.183188137332988\\
						0.6			0.189903038843253\\
						0.625		0.196482683620856\\
						0.65		0.202921293261298\\
						0.675		0.209212474917385\\
						0.7			0.215349088214072\\
						0.725		0.221323071363269\\
						0.75		0.227125201652496\\
						0.775		0.232744757832798\\
						0.8			0.238169025873206\\
						0.825		0.243382541324883\\
						0.85		0.248365857363102\\
						0.875		0.253093377680438\\
						0.9			0.257529097318862\\
						0.925		0.261616670966837\\
						0.95		0.265247718208304\\
						0.975		0.267659426204366\\
						1			0.267659426173362\\};
					\addlegendentry{$\theta_{1,\epsilon,\fix}^*$}
				\end{axis} 		
			\end{tikzpicture}
			\caption{Optimal error exponents under expected and maximum rate constraints  
				for $\epsilon:=\epsilon_1=\epsilon_2=0.05$.}\vspace{-0.3cm}
			\label{fig:DSBS_Cascaded_VL_vs_FL} 
		\end{figure}
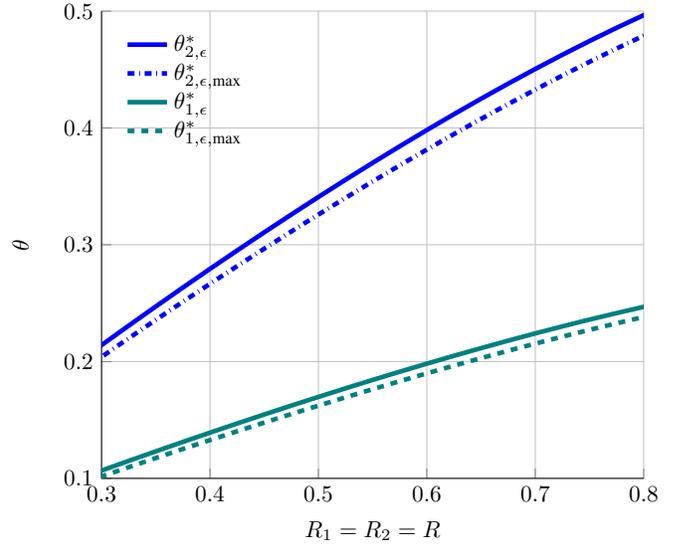

We first consider the case of equal permissible type-I error exponents $\epsilon_1=\epsilon_2$.  By Theorem~\ref{thm1}, in this case the optimal exponents region $\mathcal{E}^*$ is given by the rectangle determined by $\theta_{1,\epsilon_1}^*(R_1)$ and $ \theta_{2,\epsilon_2}^*(R_1,R_2)$. Under  maximum rate-constraints, the optimal exponents region $\mathcal{E}_{\fix}$ is also a rectangle, but now determined by $\theta_{1,\epsilon_1,\fix}^*(R_1)$ and $\theta_{2,\epsilon_2,\fix}^*(R_1,R_2)$. Fig.~\ref{fig:DSBS_Cascaded_VL_vs_FL} plots these optimal error exponents for  $\epsilon_1=\epsilon_2=0.05$ and in function of $R_1=R_2$. It thus illustrates the gain of having  expected rate constraints  instead of maximum rate-constraints.
		
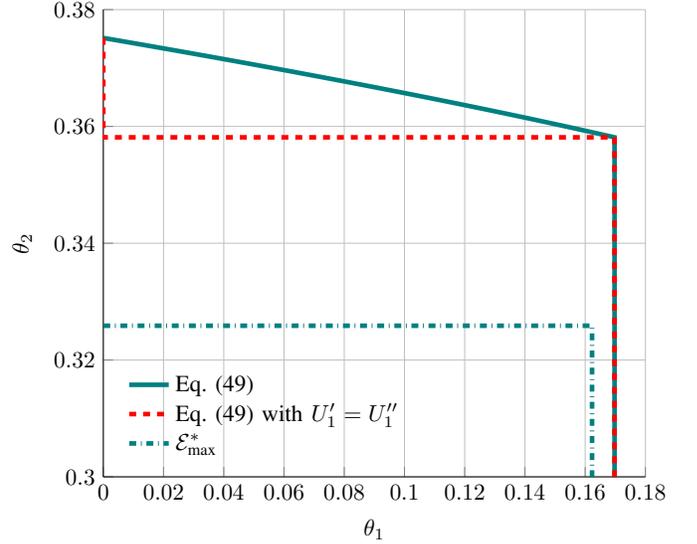
\begin{figure}[htbp]
	\begin{tikzpicture} [every pin/.style={fill=white},scale=.875]
		\begin{axis}[scale=1,
			width=.93\columnwidth,
			scale only axis,
			xmin=0,
			xmax=0.18,
			xmajorgrids,
			xlabel={$\theta_{1}$},
			x tick label style={
				/pgf/number format/.cd,
				fixed,
				precision=2,
				/tikz/.cd
			},
			ymin=0.3,
			ymax=0.38,
			ymajorgrids,
			ylabel={$\theta_2$},
			axis x line*=bottom,
			axis y line*=left,
			legend pos=south west,
			legend style={draw=none,fill=none,legend cell align=left, font=\normalsize}
			]

			\addplot[color=teal,solid,line width=2pt]
			table[row sep=crcr]{											
				0					0.375149407228070\\
				9.65894031423886e-14	0.375149407228070\\
				0.00711536871724117	0.374517176038806\\
				0.0141862538100548	0.373883773904402\\
				0.0212118435729523	0.373249205160872\\
				0.0281913021116575	0.372613474108246\\
				0.0351237681785439	0.371976585011117\\
				0.0420083539204619	0.371338542099168\\
				0.0488441435296696	0.370699349567698\\
				0.0556301917873212	0.370059011578125\\
				0.0623655224874962	0.369417532258488\\
				0.0690491267280540	0.368774915703936\\
				0.0756799610525716	0.368131165977200\\
				0.0822569454252817	0.367486287109063\\
				0.0887789610181113	0.366840283098822\\
				0.0952448477856276	0.366193157914727\\
				0.101653401799731	0.365544915494427\\
				0.108003372311205	0.364895559745400\\
				0.114293458499529	0.364245094545369\\
				0.120522305865461	0.363593523742720\\
				0.126688502212481	0.362940851156903\\
				0.132790573152924	0.362287080578831\\
				0.138826977061905	0.361632215771266\\
				0.144796099386432	0.360976260469201\\
				0.150696246197410	0.360319218380234\\
				0.156525636847444	0.359661093184933\\
				0.162282395565877	0.359001888537195\\
				0.167964541750087	0.358341608032599\\
				0.169743069578874    0.358132875286663\\
				0.169743069578874	0\\};
			\addlegendentry{Eq. \eqref{eq:E2}}
			
			\addplot[color=red,dashed,line width=2pt]
			table[row sep=crcr]{
				0  	 0.375149407228070\\
				0  	 0.358132875286663\\
				0.169743069578874    0.358132875286663\\
				0.169743069578874    0\\};
			\addlegendentry{Eq. \eqref{eq:E2} with $U_1'=U_1''$}
			
			\addplot[color=teal,dashdotted,line width=2pt]
			table[row sep=crcr]{
				0  	 0.325872480392762\\
				0.162282395565877   	0.325872480392762\\
				0.162282395565877    0\\};
			\addlegendentry{$\mathcal{E}^*_{\fix}$}
			
		\end{axis}				
	\end{tikzpicture}
	\caption{Exponents regions for   $\epsilon_1=0.05<\epsilon_2=0.15$ and $R_1=R_2=0.5$.}
	\label{fig:DSBS_Proposed_scheme_vs_FL2} 
\end{figure}

		We now consider the case $\epsilon_1=0.05< \epsilon_2=0.15$, and plot our inner bound to $\mathcal{E}^*$ in Fig.~\ref{fig:DSBS_Proposed_scheme_vs_FL2} for rates $R_1=R_2=0.5$. We note a tradeoff between the two exponents $\theta_1, \theta_2$, which was not present for $\epsilon_1=\epsilon_2$. (This tradeoff occurs because both exponents have to be optimized over the same choices of random variables $U_1',U_1''$.) The figure also shows a suboptimal version of the inner bound in Theorem~\ref{thm1}, where we set $U_1'=U_1''$ but still optimize over all choices of $U_1'$. We observe that using two different auxiliary random variables $U_1'$ and $U_1''$ (i.e., two different versions of the basic two-hop scheme) allows to obtain a better tradeoff between the two exponents. 
		Finally, for comparison,  Fig.~\ref{fig:DSBS_Proposed_scheme_vs_FL2}  also shows the exponents region $\mathcal{E}^*$ under maximum rate-constraints, so as to illustrate the gain provided by having the weaker {expected} rate constraints instead of a maximum rate constraint. 

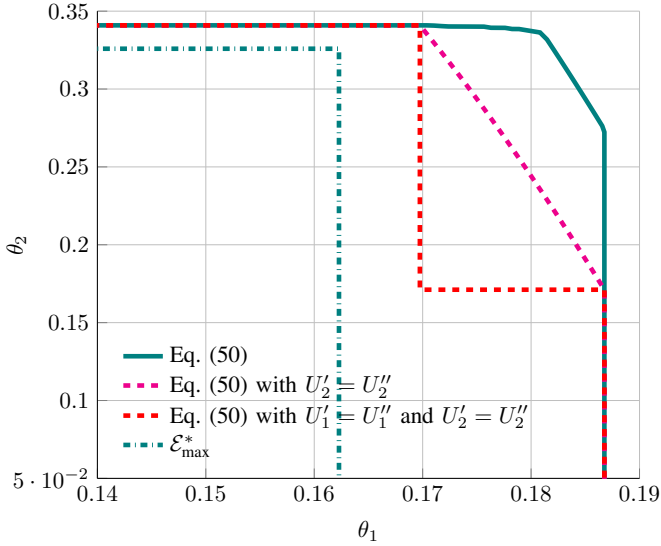
\begin{figure}[htbp]
	\begin{tikzpicture} [every pin/.style={fill=white},scale=.875]
		\begin{axis}[scale=1,
			width=.93\columnwidth,
			scale only axis,
			xmin=0.14,
			xmax=0.19,
			xmajorgrids,
			xlabel={$\theta_{1}$},
			ymin=0.05,
			ymax=0.35,
			ymajorgrids,
			ylabel={$\theta_2$},
			axis x line*=bottom,
			axis y line*=left,
			legend pos=south west,
			legend style={draw=none,fill=none,legend cell align=left, font=\normalsize}
			]
			
			\addplot[color=teal,solid,line width=2pt]
			table[row sep=crcr]{												
				0					0.340885797698502\\
				0.169743069706874	0.340885797698502\\
				0.170403689164158	0.340876366858099\\
				0.171063233936723	0.340722577355750\\
				0.171721700405229	0.340411490391450\\
				0.172379084924128	0.340161193583547\\
				0.173035383821312	0.340142274864155\\
				0.173690593397760	0.340121243244965\\
				0.174344709927168	0.340098098243934\\
				0.174997729655584	0.340072839330183\\
				0.175649648801028	0.340045465923868\\
				0.176300463553102	0.339334666549005\\
				0.176950170072602	0.339302057795125\\
				0.177598764491114	0.339267335634734\\
				0.178246242910601	0.338545189364651\\
				0.178892601402988	0.338505245890200\\
				0.179537836009735	0.337774912535338\\
				0.180181942741397	0.337039519337324\\
				0.180824917577184	0.336299086100763\\
				0.181466756464502	0.331800007038230\\
				0.182107455318493	0.325029867874832\\
				0.182747010021555	0.318209682510148\\
				0.183385416422864	0.311340405772141\\
				0.184022670337875	0.304422962593512\\
				0.184658767547818	0.297458249504565\\
				0.185293703799182	0.290447136014997\\
				0.185927474803185	0.283390465895617\\
				0.186560076235237	0.276289058369637\\
				0.186759601321881	0.272352262830495\\
				0.186759601321881	0\\};
			\addlegendentry{Eq. \eqref{eq:E3}}

			\addplot[color=magenta,dashed,line width=2pt]
			table[row sep=crcr]{													
				0					0.340885797570502\\
				0.169743069706874	0.340885797570502\\
				0.170403689164158	0.335227642938375\\
				0.171063233936723	0.329494226837738\\
				0.171721700405229	0.323687565181861\\
				0.172379084924128	0.317809565723856\\
				0.173035383821312	0.311862038186362\\
				0.173690593397760	0.305846703043338\\
				0.174344709927168	0.299765199177939\\
				0.174997729655584	0.293619090596745\\
				0.175649648801028	0.287409872346545\\
				0.176300463553102	0.281138975753178\\
				0.176950170072602	0.274807773080806\\
				0.177598764491114	0.268417581693086\\
				0.178246242910601	0.261969667784148\\
				0.178892601402988	0.255465249736315\\
				0.179537836009735	0.248905501152524\\
				0.180181942741397	0.242291553604088\\
				0.180824917577184	0.235624499128371\\
				0.181466756464502	0.228905392505933\\
				0.182107455318493	0.222135253342534\\
				0.182747010021555	0.215315067977851\\
				0.183385416422864	0.208445791239844\\
				0.184022670337875	0.201528348061215\\
				0.184658767547818	0.194563634972267\\
				0.185293703799182	0.187552521482699\\
				0.185927474803185	0.180495851363320\\
				0.186560076235237	0.173394443837340\\
				0.186759600912282    0.171142727863628\\
				0.186759600912282	0\\};
			\addlegendentry{Eq. \eqref{eq:E3} with  $U_2'=U_2''$}
			
			\addplot[color=red,dashed,line width=2pt]
			table[row sep=crcr]{
				0  	 0.340885797570502\\
				0.169743069706874   	0.340885797570502\\
				0.169743069706874   	0.171142727863628\\
				0.186759600912282    0.171142727863628\\
				0.186759600912282	0\\};
			\addlegendentry{Eq. \eqref{eq:E3} with $U_1'=U_1''$ and $U_2'=U_2''$}
			
			\addplot[color=teal,dashdotted,line width=2pt]
			table[row sep=crcr]{
				0  	 0.325872480392762\\
				0.162282395565877   	0.325872480392762\\
				0.162282395565877    0\\};
			\addlegendentry{$\mathcal{E}^*_{\fix}$}
						
		\end{axis}				
	\end{tikzpicture}
	\caption{Exponents regions under  expected and maximum rate constraints for $\epsilon_1=0.15> \epsilon_2=0.05$ and $R_1=R_2=0.5$.}
	\label{fig:DSBS_Proposed_scheme_vs_FL} 
\end{figure}

We finally consider the case $\epsilon_1=0.15>\epsilon_2=0.05$.
Fig.~\ref{fig:DSBS_Proposed_scheme_vs_FL} shows our inner bound in Theorem~\ref{thm1} together with sub-optimal versions of this inner bound where we  either set $U_2'=U_2''$ or $U_1'=U_1''$. Similarly to the previous figure we observe that having multiple auxiliary random variables (i.e., two versions of the basic two-hop scheme) allows to improve the tradeoff between the two exponents. 

\section{Conclusion}
In this work, distributed hypothesis testing over a  two-hop network with two decision centers is studied under \emph{expected} rate constraints. Different coding and decision schemes are proposed for different cases of permissible type-I error probabilities. These schemes are designed to choose different set of parameters and rates based on the transmitter's observation, aiming to maximize the achievable type-II error exponents at both decision centers. Optimal error exponents are obtained when the decision centers share equal type-I error constraints. Otherwise, a tradeoff between the exponents at the two decision centers occur. Supported by numerical simulations, the benefits of the proposed schemes are shown in this work, where the gain induced by expected rate constraints instead of maximum rate constraints is highlighted too. 

	\section*{Acknowledgment}
	M. Wigger and M. Hamad have been supported by the European Union’s Horizon 2020 Research And Innovation Programme under grant agreement no. 715111.

\vspace{-2mm}

	\bibliographystyle{ieeetr}
	\bibliography{references}
\vspace{-2mm}
\appendices

\section{Analysis of the coding scheme in Subsection~\ref{sec:scheme_same} for $\epsilon_1=\epsilon_2$}\label{app1}

Denote by $\tilde{\mathcal H}_Y$ and $\tilde{\mathcal H}_Z$ the guesses produced by the basic two-hop scheme in Subsection~\ref{sec:scheme_basic} for the chosen parameters $\mu, P_{U_1|X}, P_{U_2|Y}$. We can then write
for the type-I error probabilities:
\begin{IEEEeqnarray}{rCl}
\alpha_{1,n} &=&\Pr[\hat{\mathcal{H}}_Y=1| \mathcal{H}=0]\\
&=& \Pr[\hat{\mathcal{H}_Y} = 1, X^n \in \mathcal{S}_{n} |\mathcal{H} = 0]\nonumber \\
	&& + \Pr[\hat{\mathcal{H}}_Y = 1, X^n \notin \mathcal{S}_{n}|\mathcal{H} = 0]\IEEEeqnarraynumspace\\
	&=& \Pr[X^n \in \mathcal{S}_{n}|\mathcal{H}=0] \nonumber\\
	&& +  \Pr[\tilde{\mathcal{H}}_Y = 1, X^n \notin \mathcal{S}_{n}|\mathcal{H}=0] \IEEEeqnarraynumspace\\
	&\leq& \epsilon_1 - \mu + \Pr[\tilde{\mathcal{H}}_Y = 1|\mathcal{H}=0], 
	\end{IEEEeqnarray}
	and 
	\begin{IEEEeqnarray}{rCl}
\alpha_{2,n} &=& \Pr[\hat{\mathcal{H}}_Z=1|\mathcal{H}=0]\\
	&=& \Pr[\hat{\mathcal{H}}_Z = 1, X^n \in \mathcal{S}_{n} |\mathcal{H} = 0] \nonumber \\ 
	&& +  \Pr[\hat{\mathcal{H}}_Z = 1, X^n \notin \mathcal{S}_{n}|\mathcal{H} = 0]\IEEEeqnarraynumspace\\
&=& \Pr[X^n \in \mathcal{S}_{n}|\mathcal{H}=0] \nonumber \\ 
	&& + \Pr[\tilde{\mathcal{H}}_Z = 1, X^n \notin \mathcal{S}_{n}|\mathcal{H}=0] \IEEEeqnarraynumspace\\
	&\leq& \epsilon_2 - \mu + \Pr[\tilde{\mathcal{H}}_Z = 1|\mathcal{H}=0].
\end{IEEEeqnarray}

Since by \cite{Michele}, $\Pr[\tilde{\mathcal{H}}_Y = 1|\mathcal{H}=0]$ and $\Pr[\tilde{\mathcal{H}}_Z = 1|\mathcal{H}=0]$ both tend to 0 as $n \to  \infty$, we conclude that $\varlimsup_{n\to\infty}\alpha_{1,n} \leq \epsilon_1$, and $\varlimsup_{n\to\infty}\alpha_{2,n} \leq \epsilon_2$.

We notice that when $X^n \in \mathcal{S}_n$, then  $\hat{\mathcal H}_Y=\hat{\mathcal H}_Z=1$. The type-II error probabilities of the scheme  can therefore be bounded as:
\begin{IEEEeqnarray}{rCl}
	\beta_{1,n} &=& \Pr[\hat{\mathcal{H}}_Y=0|\mathcal{H}=1]\\
	&=& \Pr[\tilde{\mathcal{H}}_Y=0, X^n\notin \mathcal{S}_{n} |\mathcal{H}=1]\\
	&\leq& \Pr[\tilde{\mathcal{H}}_Y=0|\mathcal{H}=1]\\
	&\leq& 2^{-n\left(I(U_1;Y)+\delta(\mu)\right)} \label{BetaZLAPP1}
	\end{IEEEeqnarray}
	and 
	\begin{IEEEeqnarray}{rCl}
	\beta_{2,n} &=& \Pr[\hat{\mathcal{H}}_Z=0|\mathcal{H}=1]\\
	&=& \Pr[\tilde{\mathcal{H}}_Z=0, X^n\in \mathcal{S}_n |\mathcal{H}=1] \\
	&\leq& \Pr[\tilde{\mathcal{H}}_Z=0|\mathcal{H}=1]\IEEEeqnarraynumspace\\
	&\leq& 2^{-n\left(I(U_1;Y)+ I(U_2;Z)+\delta(\mu)\right)} \IEEEeqnarraynumspace  \label{BetaSWWAPP1}
\end{IEEEeqnarray}
where \eqref{BetaZLAPP1} and \eqref{BetaSWWAPP1} are proved in \cite{Michele}, and $\delta(\mu) \to 0$ as $\mu \downarrow 0$.

The described scheme satisfies the rate constraints for all blocklengths $n$ that are sufficiently large so that  $(1-\epsilon_1+\mu)n\mu  \geq (\epsilon_{1} - \mu) \; \Leftrightarrow \; (1-\epsilon_2+\mu)n\mu\geq  (\epsilon_{2} - \mu)$ hold:
\begin{IEEEeqnarray}{rCl}
	\E({\len(\M_1)})& \leq& (\epsilon_1-\mu) + (1-\epsilon_1 +\mu)\cdot n( I(U_1;X) + \mu) \IEEEeqnarraynumspace\\
	&\leq  & (1-\epsilon_1 +\mu)\cdot n( I(U_1;X) + 2\mu) \label{eq:n_large_enough_1}\\
	&=  & n R_1
		\end{IEEEeqnarray}
	and 
	\begin{IEEEeqnarray}{rCl}
		\E({\len(\M_2)})& \leq &(\epsilon_2-\mu)+(1-\epsilon_2 +\mu)\cdot n( I(U_2;Y) + \mu)\IEEEeqnarraynumspace\\
			&\leq  & (1-\epsilon_2 +\mu)\cdot n( I(U_2;Y) + 2\mu) \label{eq:n_large_enough_2}\\
	&=& n R_2.
	\end{IEEEeqnarray}
Letting first $n \to \infty$ and then  $\mu \downarrow 0$, establishes the desired achievability result in \eqref{eq:E1}.

\section{Analysis of the coding scheme in Subsection~\ref{sec:scheme_larger} for  $\epsilon_2>\epsilon_1$}\label{app2}
Let  ${\tilde{\mathcal{H}}'}_Y$ and  $\tilde{\mathcal{H}}'_Z$ denote the hypotheses guessed by {\Rel} and {\Rec} for the basic two-hop scheme with the first parameter choices $\mu, P_{U_1'|X}, P_{U_2'|Y}$. Similarly, let ${\tilde{\mathcal{H}}''}_Y$ be the hypothesis  produced by {\Rel} for the basic two-hop scheme with the parameter choice $\mu, P_{U_1''|X}$. 
We then obtain for the type-I error probabilities: 
\begin{IEEEeqnarray}{rCl}
	\alpha_{1,n} &=& \Pr[\hat{\mathcal{H}}_Y = 1, X^n \in \mathcal{S}_{n} |\mathcal{H} = 0] \nonumber \\ 
	&& + \Pr[\hat{\mathcal{H}}_Y = 1, X^n \in \mathcal{D}_{n}' |\mathcal{H} = 0] \nonumber \\ 
	&& +  \Pr[\hat{\mathcal{H}}_Y = 1, X^n \in \mathcal{D}_n''|\mathcal{H} = 0]\IEEEeqnarraynumspace\\
	&=& \Pr[X^n \in \mathcal{S}_{n}|\mathcal{H}=0] \nonumber \\ 
	&& +  \Pr[\tilde{\mathcal{H}}'_Y = 1, X^n \in \mathcal{D}_n'|\mathcal{H}=0] \nonumber \\
	&& + \Pr[\tilde{\mathcal{H}}''_Y = 1, X^n \in \mathcal{D}_{n}''|\mathcal{H}=0] \IEEEeqnarraynumspace\\
	&\leq& \epsilon_1 - \mu + \Pr[\tilde{\mathcal{H}}'_Y = 1|\mathcal{H}=0] \nonumber\\ && + \Pr[\tilde{\mathcal{H}}''_Y = 1|\mathcal{H}=0]
\end{IEEEeqnarray}
and 
\begin{IEEEeqnarray}{rCl}
	\alpha_{2,n} &=& \Pr[\hat{\mathcal{H}}_Z = 1, X^n \in (\mathcal{S}_{n}\cup\mathcal{D}_n'') |\mathcal{H} = 0] \nonumber \\
	&& + \Pr[\hat{\mathcal{H}}_Z = 1, X^n \in \mathcal{D}_{n}'|\mathcal{H} = 0]\IEEEeqnarraynumspace\\
	&=& \Pr[X^n \in (\mathcal{S}_{n}\cup\mathcal{D}_n'')|\mathcal{H}=0] \nonumber\\
	&& +  \Pr[\tilde{\mathcal{H}}'_Z = 1, X^n \in \mathcal{D}_{n}'|\mathcal{H}=0] \IEEEeqnarraynumspace\\
	&\leq& \epsilon_2 - \mu + \Pr[\tilde{\mathcal{H}}'_Z = 1|\mathcal{H}=0].
\end{IEEEeqnarray}
Since by \cite{Michele},  $\Pr[\tilde{\mathcal{H}}'_Y = 1|\mathcal{H}=0]$ and $ \Pr[\tilde{\mathcal{H}}''_Y = 1|\mathcal{H}=0], \textnormal{and} \Pr[\tilde{\mathcal{H}}'_Z = 1|\mathcal{H}=0] \downarrow 0$ as $n \to \infty$, we conclude that for the  scheme in Subsection~\ref{sec:scheme_larger} $\varlimsup_{n\to\infty}\alpha_{1,n} \leq \epsilon_1$ and $\varlimsup_{n\to\infty}\alpha_{2,n} \leq \epsilon_2$.

For the type-II error probabilities we obtain
\begin{IEEEeqnarray}{rCl}
	\beta_{1,n} 
	&=& \Pr[\tilde{\mathcal{H}}'_Y=0, X^n\in \mathcal{D}_n' |\mathcal{H}=1] \nonumber \\
	&& + \Pr[\tilde{\mathcal{H}}''_Y=0, X^n\in \mathcal{D}_{n}''|\mathcal{H}=1]\\
	&\leq& \Pr[\tilde{\mathcal{H}}'_Y=0|\mathcal{H}=1] + \Pr[\tilde{\mathcal{H}}''_Y=0|\mathcal{H}=1]\IEEEeqnarraynumspace\\
	&\leq& 2^{-n(I(U_1';Y) + \delta(\mu))} + 2^{-n(I(U_1'';Y) + \delta(\mu))} \IEEEeqnarraynumspace  \label{BetaSWWAPP2}
\end{IEEEeqnarray}\vspace{-2mm}
and 
\begin{IEEEeqnarray}{rCl}
	\beta_{2,n} 
	&=& \Pr[\tilde{\mathcal{H}}'_Z=0, X^n\in \mathcal{D}_{n}' |\mathcal{H}=1]\\
	&\leq& \Pr[\tilde{\mathcal{H}}'_Z=0|\mathcal{H}=1]\\
	&\leq& 2^{-n\left(I(U_1';Y)+I(U_2';Z) + \delta(\mu)\right)}, \label{BetaZLAPP2}
\end{IEEEeqnarray}
where \eqref{BetaSWWAPP2} and \eqref{BetaZLAPP2} are proved in \cite{Michele}, and $\delta(\mu) \downarrow 0$ as $\mu \downarrow 0$. 

The described scheme satisfies the rate constraints for all blocklengths $n$ that are sufficiently large so that both $(1-\epsilon_1+\mu)n\mu \geq (2 - \epsilon_1+\mu)$ and $(1-\epsilon_2+\mu)n\mu \geq (2-\epsilon_1+\mu)$	 hold:
\begin{IEEEeqnarray}{rCl}
	\E[\len(\M_1)]& \leq& (\epsilon_1-\mu) \nonumber \\ && +(1-\epsilon_2 +\mu)\cdot (n( I(U_1';X) + \mu)+ 2) \nonumber \\
	&& + ( \epsilon_2-\epsilon_1  )\cdot (n(I(U_1'';X) + \mu) +2)\\
	& \leq & n(R_1'+R_1'')=n R_1 \label{eq:n_large_enough_shortcut_1}
\end{IEEEeqnarray}
and 
\begin{IEEEeqnarray}{rCl}
	\E[\len(\M_2)]& \leq & (\epsilon_1-\mu) + (\epsilon_{2}-\epsilon_1) \cdot 2\nonumber \\
	&&+(1-\epsilon_2 +\mu)\cdot (n( I(U_2';Y) + \mu)+2) \IEEEeqnarraynumspace\\
	& \leq & nR_2. \label{eq:n_large_enough_shortcut_2}
\end{IEEEeqnarray}
Letting first $n \to \infty$ and then  $\mu \downarrow 0$, establishes the desired result in \eqref{eq:E2}. 

\section{Analysis of the coding scheme in Subsection~\ref{sec:scheme1_larger} for  $\epsilon_1>\epsilon_2$}\label{app3}
 Let  ${\tilde{\mathcal{H}}'}_Y$ and  $\tilde{\mathcal{H}}'_Z$ denote the hypotheses guessed by {\Rel} and {\Rec} for the basic two-hop scheme with the first parameter choices $\mu, P_{U_1'|X}, P_{U_2'|Y}$. Similarly, let ${\tilde{\mathcal{H}}''}_Z$ be the hypothesis  produced by {\Rec} for the basic two-hop scheme with the parameter choices $\mu, P_{U_1''|X}, P_{U_2''|Y}$. 
We then obtain for the type-I error probabilities:  
	\begin{IEEEeqnarray}{rCl}
	\alpha_{1,n} &=& \Pr[\hat{\mathcal{H}}_Y = 1, X^n \in (\mathcal{S}_{n}\cup\mathcal{D}_n'') |\mathcal{H} = 0] \nonumber \\
	&& + \Pr[\hat{\mathcal{H}}_Y = 1, X^n \in \mathcal{D}_{n}'|\mathcal{H} = 0]\IEEEeqnarraynumspace\\
	&=& \Pr[X^n \in (\mathcal{S}_{n}\cup\mathcal{D}_n'')|\mathcal{H}=0] \nonumber\\
	&& +  \Pr[\tilde{\mathcal{H}}'_Y = 1, X^n \in \mathcal{D}_{n}'|\mathcal{H}=0] \IEEEeqnarraynumspace\\
	&\leq& \epsilon_1 - \mu + \Pr[\tilde{\mathcal{H}}'_Y = 1|\mathcal{H}=0]
\end{IEEEeqnarray}
and
\begin{IEEEeqnarray}{rCl}
	\alpha_{2,n} &=& \Pr[\hat{\mathcal{H}}_Z = 1, X^n \in \mathcal{S}_{n} |\mathcal{H} = 0] \nonumber \\ 
	&& + \Pr[\hat{\mathcal{H}}_Z = 1, X^n \in \mathcal{D}_{n}' |\mathcal{H} = 0] \nonumber \\ 
	&& +  \Pr[\hat{\mathcal{H}}_Z = 1, X^n \in \mathcal{D}_n''|\mathcal{H} = 0]\IEEEeqnarraynumspace\\
	&=& \Pr[X^n \in \mathcal{S}_{n}|\mathcal{H}=0] \nonumber \\ 
	&& +  \Pr[\tilde{\mathcal{H}}'_Z = 1, X^n \in \mathcal{D}_n'|\mathcal{H}=0] \nonumber \\
	&& + \Pr[\tilde{\mathcal{H}}''_Z = 1, X^n \in \mathcal{D}_{n}''|\mathcal{H}=0] \IEEEeqnarraynumspace\\
	&\leq& \epsilon_2 - \mu + \Pr[\tilde{\mathcal{H}}'_Z = 1|\mathcal{H}=0] \nonumber\\ && + \Pr[\tilde{\mathcal{H}}''_Z = 1|\mathcal{H}=0].
\end{IEEEeqnarray}
Since by \cite{Michele},  $\Pr[\tilde{\mathcal{H}}'_Y = 1|\mathcal{H}=0]$, $ \Pr[\tilde{\mathcal{H}}'_Z = 1|\mathcal{H}=0], \textnormal{and} \Pr[\tilde{\mathcal{H}}''_Z = 1|\mathcal{H}=0]$ all tend to 0 as $n \to \infty$, we conclude that for the  scheme in Subsection~\ref{sec:scheme1_larger} $\varlimsup_{n\to\infty}\alpha_{1,n} \leq \epsilon_1$ and $\varlimsup_{n\to\infty}\alpha_{2,n} \leq \epsilon_2$.

For the type-II error probabilities we obtain
\begin{IEEEeqnarray}{rCl}
	\beta_{1,n} 
	&=& \Pr[\tilde{\mathcal{H}}'_Y=0, X^n\in \mathcal{D}_{n}' |\mathcal{H}=1]\\
	&\leq& \Pr[\tilde{\mathcal{H}}'_Y=0|\mathcal{H}=1]\\
	&\leq& 2^{-n\left(I(U_1';Y)+ \delta(\mu)\right)}, \label{BetaZLAPP3}
\end{IEEEeqnarray}\vspace{-2mm}
and 
\begin{IEEEeqnarray}{rCl}
	\beta_{2,n} 
	&=& \Pr[\tilde{\mathcal{H}}'_Z=0, X^n\in \mathcal{D}_n' |\mathcal{H}=1] \nonumber \\
	&& + \Pr[\tilde{\mathcal{H}}''_Z=0, X^n\in \mathcal{D}_{n}''|\mathcal{H}=1]\\
	&\leq& \Pr[\tilde{\mathcal{H}}'_Z=0|\mathcal{H}=1] + \Pr[\tilde{\mathcal{H}}''_Z=0|\mathcal{H}=1]\IEEEeqnarraynumspace\\
	&\leq& 2^{-n\left(I(U_1';Y)+I(U_2';Z) + \delta(\mu)\right)} \nonumber \\ 
	&& + 2^{-n\left(I(U_1'';Y)+I(U_2'';Z) + \delta(\mu)\right)}. \IEEEeqnarraynumspace  \label{BetaSWWAPP3}
\end{IEEEeqnarray}
where \eqref{BetaSWWAPP3} and \eqref{BetaZLAPP3} are proved in \cite{Michele}, and $\delta(\mu) \downarrow 0$ as $\mu \downarrow 0$. 

The described scheme satisfies the rate constraints for all blocklengths $n$ that are sufficiently large so that $(1-\epsilon_2+\mu)n\mu \geq ( 2 - \epsilon_{2} +  \mu)$	 holds:	
	\begin{IEEEeqnarray}{rCl}
	\E[\len(\M_1)]& \leq& (\epsilon_2-\mu) \nonumber \\ && +(1-\epsilon_1 +\mu)\cdot (n( I(U_1';X) + \mu)+2) \nonumber \\
	&& + ( \epsilon_1-\epsilon_2  )\cdot (n(I(U_1'';X) + \mu) +2)\\
	& \leq & n (R_1' + R_1'') = nR_1 \label{eq:n_large_enough_case3}
	\end{IEEEeqnarray}
	and 
	\begin{IEEEeqnarray}{rCl}
	\E[\len(\M_2)]& \leq & (\epsilon_2-\mu)\nonumber \\
	&&+(1-\epsilon_1 +\mu)\cdot (n( I(U_2';Y) + \mu)+2) \nonumber \\
	&& + ( \epsilon_1-\epsilon_2  )\cdot (n(I(U_2'';Y) + \mu)  +2) \label{eq:inequality_length_M2}\\
	& \leq & n(R_2'+R_2'') = nR_2. \label{eq:n_large_enough_case3_2}
\end{IEEEeqnarray}
Letting first $n \to \infty$ and then  $\mu \downarrow 0$, establishes the desired result in \eqref{eq:E3}.

\section{Converse Proof to Theorem~\ref{thm1} when $\epsilon_{1}=\epsilon_{2}$}\label{converse}
Throughout this section, let $h_{b}(\cdot)$ denote the binary entropy function, and $D(P\|Q)$ denote the Kullback-Leibler divergence between two probability mass functions on the same alphabet. 

Define $\epsilon \triangleq \epsilon_{1} = \epsilon_{2}$ and fix $\theta_{1} < \theta_{1,\epsilon}^*(R_1)$ and $\theta_{2} < \theta_{2,\epsilon}^*(R_1,R_2)$. 
The proof consists of three parts. In the first two parts (Subsections~\ref{sec:part1}-\ref{sec:part2}) we establish constraints based on the decisions at {\Rel} and at {\Rec}, respectively, and in the third part we combine the constraints. 

\subsection{Constraints based on \Rel's decision}\label{sec:part1}
Considering only the decision at {\Rel} but not  at \Rec, by  \cite{JSAIT} we conclude that there exists an auxiliary random variable ${U}_1'$ jointly distributed with the pair $(X,Y)\sim P_{XY}$ so that the following conditions hold:
\begin{IEEEeqnarray}{rCl}
	\theta_{1} &\leq&  I({U}_1';Y),\label{eq:thetaR_ub1}\\
	R_1 &\geq& (1-\epsilon)I({U}_1';X),\label{eq:R1_lb1}\\
	{U}_1' &\to& X \to Y. \label{eq:MC_U_1}
\end{IEEEeqnarray}

\subsection{Constraints based on \Rec's decision}\label{sec:part2}
In what follows we establish similar constraints but based on the decision at \Rec. The proof is basically an extension of the proof in \cite{JSAIT} but to a multi-hop network. Consider  a sequence  of encoding and decision functions  $\{(\phi_1^{(n)}, \phi_2^{(n)}, g_2^{(n)})\}_{n \geq 1}$ satisfying the type-I and type-II error constraints \eqref{type1constraint1}--\eqref{thetaconstraint} for $i=2$. Then, fix  a blocklength $n$ and a small number $\eta \geq 0$ and define
\begin{IEEEeqnarray}{rCl}\label{Bn}
	\mu_n &\triangleq& n^{-{1\over3}},\\
	\mathcal{B}_n(\eta) &\triangleq& \{(x^n,y^n) : \nonumber\\ && \; \mathrm{Pr}[\hat{\mathcal{H}}_Z=0 | X^n = x^n, Y^n = y^n, \mathcal{H}=0] \geq \eta\}, \IEEEeqnarraynumspace\label{eq:Bndef}\\
	\mathcal{D}_n(\eta) &\triangleq& \mathcal{B}_n(\eta) \cap \mathcal{T}_{\mu_n}^{n}(P_{XY}). \label{Dn2}   
\end{IEEEeqnarray}
By constraint (\ref{type1constraint1}) on the type-I error probability, we have:
\begin{IEEEeqnarray}{rCl}
	1 - \epsilon &\leq& \Pr[\hat{\mathcal{H}}_Z =0 | \mathcal{H}=0]\\ 
	&=& \sum_{(x^n,y^n)\in \mathcal{B}_n} \underbrace{\Pr[\hat{\mathcal{H}}_Z =0 | X^n = x^n, Y^n = y^n ,\mathcal{H}=0]}_{ \leq 1}\nonumber\\
	&&\quad\quad\;\cdot P_{X^nY^n}(x^n,y^n) \nonumber \\ &&  + \sum_{(x^n,y^n)\notin \mathcal{B}_n} \underbrace{\Pr[\hat{\mathcal{H}}_Z =0 | X^n = x^n, Y^n = y^n ,\mathcal{H}=0]}_{\leq \eta}\nonumber\\
	&&\qquad\quad\; \cdot P_{X^nY^n}(x^n,y^n) \\ 
	&\leq& P_{X^nY^n}(\mathcal{B}_n(\eta))  + \eta (1- P_{X^nY^n}(\mathcal{B}_{n}(\eta))),
\end{IEEEeqnarray}	
	and thus 
	\begin{IEEEeqnarray}{rCl}
 P_{X^nY^n}(\mathcal{B}_n(\eta)) \geq {1 - \epsilon - \eta \over{1 - \eta}} . \label{Bnprob}
\end{IEEEeqnarray}
Moreover, by \cite[Remark to Lemma~2.12]{Csiszarbook}, the probability that the pair ($X^n,Y^n$) lies in the jointly strong typical set $\mathcal{T}_{\mu_n}^{(n)}(P_{XY})$ satisfies
\begin{equation}\label{Txy}
	P_{XY}^{n}\left(\mathcal{T}_{\mu_n}^{(n)}(P_{XY})\right) \geq 1 - {\vert{\mathcal{X}}\vert\ \vert{\mathcal{Y}}\vert\ \over{4 \mu_n^2 n}},
\end{equation}
and thus by \eqref{Dn2} and \eqref{Bnprob}, 
\begin{equation}\label{Dn2prob}
	P_{XY}^n(\mathcal{D}_n) \geq {1-\epsilon - \eta \over {1 - \eta}} - {\vert{\mathcal{X}}\vert\ \vert{\mathcal{Y}}\vert\ \over{4 \mu_n^2 n}} \triangleq \Delta_n. 
\end{equation}
We define the random variables $\left({\tilde{\M}}_1,{\tilde{\M}_2},\tilde{X}^n,{\tilde{Y}^n},\tilde{Z}^n\right)$ as the restriction of the random variables $({\M_1},{\M_2},{X^n},{Y^n},{Z^n})$ to $(X^n,Y^n) \in D_n(\eta)$ with their probability distribution given by:
\begin{IEEEeqnarray}{rCl}
	\lefteqn{P_{{\tilde{\M}_1}{\tilde{\M}_2}\tilde{X}^n\tilde{Y}^n\tilde{Z}^n}(\m_1,\m_2,x^n,y^n,z^n) \triangleq} \qquad \nonumber \\
	&& P_{X^nY^nZ^n}(x^n,y^n,z^n)\cdot{\mathbbm{1} \{(x^n,y^n)\in D_n(\eta)\} \over P_{X^nY^n}(D_n(\eta))}\nonumber\\ && \qquad \cdot{\mathbbm{1}\{\phi_1(x^n)=\m_1\}}
	\cdot{\mathbbm{1}\{\phi_2(y^n,\phi_1(x^n))=\m_2\}}, \IEEEeqnarraynumspace \label{pmftildedoubleprime}
\end{IEEEeqnarray}
leading to the following inequalities:
\begin{equation}\label{tildem1m2relation}
	P_{\tilde{\M}_1\tilde{\M}_2}(\m_1,\m_2) \leq P_{\M_1\M_2}(\m_1,\m_2) \Delta_{n}^{-1},
\end{equation}
\begin{equation}\label{tildeyrelation}
	P_{\tilde{Z}^n}(z^n) \leq P_{Z}^{n}(z^n) \Delta_n^{-1},
\end{equation}
\begin{equation}\label{tildedivergencerelation}
	D(P_{\tilde{X}^n\tilde{Y}^n}||P_{XY}^{n}) \leq \log{\Delta_n^{-1}}.
\end{equation}

\subsubsection{Single-Letter Characterization of Rate Constraints}

Define the following random variables:
\begin{equation}
	{\tilde{L}_i} \triangleq \mathrm{len}({\tilde{\M}_i}),\;\;\;\;\;\; i=1,2.
\end{equation}
By the rate constraints  \eqref{eq:Rate1} and \eqref{eq:Rate2}, we get under $\mathcal{H} = 0$:
\begin{IEEEeqnarray}{rCl}
	nR_i &\geq& \mathbb{E}[L_i]\\
	&\geq& \mathbb{E}[L_i|(X^n,Y^n) \in \mathcal{D}_n(\eta)]P_{X^nY^n}(\mathcal{D}_n(\eta))\\
	&=& \mathbb{E}[\tilde{L}_i]P_{X^nY^n}(\mathcal{D}_n(\eta))\\ 
	&\geq&\mathbb{E}[\tilde{L}_i]\Delta_n, \label{ELiprime}
\end{IEEEeqnarray} 
where the last inequality holds by \eqref{Dn2prob}.	
Moreover, by definition, $\tilde{L}_i$ is a function of $\tilde{\M}_i$, for $i=1,2$, so we can upper bound the entropy of $\tilde{\M}_i$ as follows:
\begin{align}
	H(\tilde{\M}_i) &= H(\tilde{\M}_i,\tilde{L}_i)\\
	&= \sum_{l_i} \Pr[\tilde{L}_i = l_i]H(\tilde{\M}_i|\tilde{L}_i=l_i) + H(\tilde{L}_i)\\
	&\leq \sum_{l_i} \Pr[\tilde{L}_i = l_i]l_i + H(\tilde{L}_i) \label{HMi_ineq1}\\
	& = \mathbb{E}[\tilde{L}_i] + H(\tilde{L}_i)\\
	&\leq {nR_i\over \Delta_n} + {nR_i\over \Delta_n}{h_b\left({\Delta_n \over nR_i}\right)} \label{HMi_ineq3}\\
	&= {nR_i\over \Delta_n} \left(1 + {h_b\left({\Delta_n \over nR_i}\right)}\right), \label{Miub}
\end{align}
where \eqref{HMi_ineq3} holds by \eqref{ELiprime} and since the maximum possible entropy of $\tilde{L}_i$ is obtained by a geometric distribution of mean $\mathbb{E}[\tilde{L}_i]$, which is further bounded by $nR_i \over \Delta_n$ \cite[Theorem 12.1.1]{cover}.

On the other hand, we lower bound the entropy of $\tilde{\M}_i$ as:
\begin{IEEEeqnarray}{rCl}
	H(\tilde{\M}_i) &\geq& I(\tilde{\M}_i;\tilde{X}^n\tilde{Y}^n) + D(P_{\tilde{X}^n\tilde{Y}^n}||P_{XY}^n) + \log\Delta_{n}\IEEEeqnarraynumspace\label{m1entropylbstep1}\\
	&=& H(\tilde{X}^n\tilde{Y}^n) + D(P_{\tilde{X}^n\tilde{Y}^n}||P_{XY}^n) \nonumber\\& &-  H(\tilde{X}^n\tilde{Y}^n|\tilde{\M}_i) + \log\Delta_{n}\\
	&\geq& n [H(\tilde{X}_{T}\tilde{Y}_{T}) + D(P_{\tilde{X}_{T}\tilde{Y}_{T}}||P_{XY})] \nonumber\\
	& &- \sum_{t=1}^{n} H(\tilde{X}_{t}\tilde{Y}_{t}|\tilde{U}_{i,t})+ \log\Delta_{n}\label{m1entropylbstep4}\\
	&=& n [H(\tilde{X}_{T}\tilde{Y}_{T}) + D(P_{\tilde{X}_{T}\tilde{Y}_{T}}||P_{XY})] \nonumber\\
	& &- n H(\tilde{X}_{T}\tilde{Y}_{T}|\tilde{U}_{i,T},T)+ \log\Delta_{n} \label{Tuniformdef}
\end{IEEEeqnarray}
Here, (\ref{m1entropylbstep1}) holds by (\ref{tildedivergencerelation});  (\ref{m1entropylbstep4}) holds by the super-additivity property in \cite[Proposition 1]{tyagi2019strong}, by the chain rule, and by defining 
\begin{equation}\label{eq:Utilde_t}
\tilde{U}_{i,t} \triangleq (\tilde{\M}_i\tilde{X}^{t-1}\tilde{Y}^{t-1}), \quad i\in\{1,2\};
\end{equation}
and  \eqref{Tuniformdef} holds  by defining $T$ uniform over $\{1,\dots,n\}$ independent of all other random variables. Finally, defining 
\begin{subequations}\label{eq:Utilde}
\begin{IEEEeqnarray}{rCl}
\tilde{U}_i &\triangleq& (\tilde{U}_{i,T},T), \quad i\in\{1,2\}, \\
 \tilde{X} &\triangleq &\tilde{X}_{T}\\
 \tilde{Y} &\triangleq & \tilde{Y}_{T}
\end{IEEEeqnarray}
\end{subequations}
results in:
\begin{IEEEeqnarray}{rCl}
	H(\tilde{\M}_1) &\geq& n \left[I(\tilde{X};\tilde{U}_1) + {1 \over n} \log{\Delta_{n}}\right], \label{eq:HM1_LB'}\\
	H(\tilde{\M}_2) &\geq& n \left[I(\tilde{Y};\tilde{U}_2) + {1 \over n} \log{\Delta_{n}}\right], \label{eq:HM2_LB}
\end{IEEEeqnarray}
thus following \eqref{Miub}, \eqref{eq:HM1_LB'}, and \eqref{eq:HM2_LB}, we deduce that:
\begin{equation}\label{R1lb}
	R_1 \geq {I(\tilde{X};\tilde{U}_1) + {1 \over n} \log \Delta_{n}  \over {1 + h_b\left({\Delta_n \over nR_1}\right)}}\cdot\Delta_n,
\end{equation}
\begin{equation}\label{R2lb}
	R_2 \geq {{I(\tilde{Y};\tilde{U}_2) + {1 \over n} \log \Delta_{n}} \over {1 + h_b\left({\Delta_n \over nR_2}\right)}}\cdot\Delta_n. 
\end{equation}

\subsubsection{Upper Bounding the Type-II Error Exponent $\theta_2$}

\qquad Define for each $\m_2$ the set
\begin{equation}
	\mathcal{A}_{Z,n}(\m_2) \triangleq \{z^n \colon g_2(\m_2,z^n) = 0\},
\end{equation}
and its \emph{Hamming neighborhood}
\begin{equation}
	\hat{\mathcal{A}}_{Z,n}^{\ell_n}(\m_2) \triangleq \{\tilde{z}^n : \exists \, z^n \in \,\mathcal{A}_{Z,n}(\m_2) \textnormal{ s.t.} \; d_H(z^n,\tilde{z}^n)\leq\ell_n\}
\end{equation}
for some real number $\ell_n$ satisfying $\lim_{n \rightarrow \infty} {\ell_n/n} =0 $ and $\lim_{n \rightarrow \infty} {\ell_n/\sqrt{n}} =\infty $.	
Note that: 
\begin{equation}
	\mathcal{A}_{Z,n} = \bigcup\limits_{\m_2 \in \mathcal{M}_2} \{\m_2\}\times\mathcal{A}_{Z,n}(\m_2).
\end{equation} 
Since by definitions (\ref{Bn}) and (\ref{Dn2}), for all $(x^n,y^n) \in \mathcal{D}_n(\eta)$, and where $\m_2 = \phi_2(\phi_1(x^n),y^n)$:
\begin{equation}\label{blowupcond2}
	P_{\tilde{Z}^n|\tilde{X}^n\tilde{Y}^n}(\mathcal{A}_{Z,n}(\m_2)|x^n,y^n) \geq \eta ,
\end{equation}
then by the blowing-up lemma \cite{MartonBU}:
\begin{equation}\label{blowup2}
	P_{\tilde{Z}^n|\tilde{X}^n\tilde{Y}^n}(\hat{\mathcal{A}}_{Z,n}^{\ell_n}(\m_2)|x^n,y^n) \geq 1 - \zeta_n,
\end{equation}
for a real number $\zeta_n > 0$ such that $\lim\limits_{n \to \infty} \zeta_n = 0$.
Therefore:
\begin{IEEEeqnarray}{rCl}
	P_{\tilde{\M}_2\tilde{Z}^n}(\hat{\mathcal{A}}_{Z,n}^{\ell_n})
	&=& \sum_{\substack{(x^n,y^n)\in\mathcal{D}_n\\\m_2\in \mathcal{M}_2}}P_{\tilde{Z}^n|\tilde{X}^n\tilde{Y}^n}(\hat{\mathcal{A}}_{Z,n}^{\ell_n}(\m_2)|x^n,y^n)\nonumber\\ 
	&& \qquad\qquad \cdot P_{\tilde{X}^n\tilde{Y}^n\tilde{\M}_2}(x^n,y^n,\m_2)\IEEEeqnarraynumspace\\
	&\geq&  (1-\zeta_n).
\end{IEEEeqnarray}
Now define:
\begin{equation}
	Q_{\tilde{\M}_2}(\m_2) \triangleq \sum_{y^n,\m_1} P_{\tilde{\M}_1}(\m_1)P_{\tilde{Y}^n}(y^n)\cdot \mathbbm{1}\{\phi_2(\m_1,y^n)=\m_2\},
\end{equation}
and 
\begin{IEEEeqnarray}{rCl}
	Q_{{\M}_2}(\m_2) &=& \sum_{x^n,y^n,z^n,\m_1}P_{X}^n(x^n)P_{Y}^n(y^n)P_{Z}^n(z^n)\nonumber\\
	&&\quad\quad \cdot\mathbbm{1}\{\phi_1(x^n)=\m_1,\phi_2(\m_1,y^n)=\m_2\} \IEEEeqnarraynumspace\\
	&=&\sum_{x^n,y^n,\m_1}P_{X^n\M_1}(x^n,\m_1)P_{Y}^n(y^n)\nonumber\\
	&&\quad\quad\cdot\mathbbm{1}\{\phi_2(\m_1,y^n)=\m_2\} \\
	&=&	\sum_{y^n,\m_1} \hspace{-1.5mm}P_{{\M}_1}(\m_1)P_{Y}^n(y^n)\cdot \mathbbm{1}\{\phi_2(\m_1,y^n)=\m_2\}\IEEEeqnarraynumspace
\end{IEEEeqnarray}
Then 
\begin{equation}
	Q_{\tilde{\M}_2}(\m_2) \leq Q_{\M_2}(\m_2)\Delta_n^{-2},
\end{equation}
and
\begin{IEEEeqnarray}{rCl}
	\lefteqn{Q_{\tilde{\M}_2}P_{\tilde{Z}^n}\left(\hat{\mathcal{A}}_{Z,n}^{\ell_n}\right)} \qquad  \nonumber\\
	&\leq& Q_{\M_2}P_{Z}^n\left(\hat{\mathcal{A}}_{Z,n}^{\ell_n}\right)\Delta_n^{-3}\\
	&\leq& \underbrace{Q_{\M_2}P_{Z}^n\left(\mathcal{A}_{Z,n}\right)}_{\beta_{2,n}}e^{nh_b(\ell_n/n)}{|\mathcal{Z}|^{\ell_n}k_n^{\ell_n}}\Delta_n^{-3}\label{Eq:ByCsiszarKornerLemma}\\
	&=& \beta_{2,n}F_n^{\ell_n}\Delta_n^{-3},
\end{IEEEeqnarray}
where $k_n \triangleq \min\limits_{\substack{z,z':P_Z(z') > 0}}{P_Z(z) \over P_Z(z')}$ and $ F_n^{\ell_n} \triangleq e^{nh_b(\ell_n/n)} \cdot k^{\ell_n} \cdot {\vert \mathcal{Z} \vert}^{\ell_n}$. Here, (\ref{Eq:ByCsiszarKornerLemma}) holds by \cite[Proof of Lemma 5.1]{Csiszarbook}.
Then by standard inequalities (see \cite[Lemma~1]{JSAIT}), we can obtain the following expression:
\begin{equation}\label{theta_ub}
	{1\over n}\log{1 \over \beta_{2,n}} \leq {1 \over n (1-\zeta_n)} \left(D(P_{\tilde{\M}_2\tilde{Z}^n}||Q_{\tilde{\M}_2}P_{\tilde{Z}^n}) + 1\right) + \delta_n
\end{equation}
where $\delta_n$ tends to 0 as $n \to \infty$. 

We further upper bound the divergence terms as follows:
	\begin{IEEEeqnarray}{rCl}
	\lefteqn{D(P_{\tilde{\M}_2\tilde{Z}^n}||Q_{\tilde{\M}_2}P_{\tilde{Z}^n})}\qquad \nonumber\\
	&=& I(\tilde{\M}_2;\tilde{Z}^n) + D(P_{\tilde{\M}_2}||Q_{\tilde{\M}_2}) \\
	&\leq& I(\tilde{\M}_2;\tilde{Z}^n) + D(P_{\tilde{Y}^n\tilde{\M}_1}||P_{\tilde{Y}^n}P_{\tilde{\M}_1})\label{eq:dp_ineq_relative_entropy}\\
	&=& I(\tilde{\M}_2;\tilde{Z}^n) + I(\tilde{\M}_1;\tilde{Y}^n)\\
	&=& \sum_{t=1}^n I(\tilde{\M}_2;\tilde{Z}_t|\tilde{Z}^{t-1}) + I(\tilde{\M}_1;\tilde{Y}_t|\tilde{Y}^{t-1})\label{eq:divergence_chainrule}\\
	&\leq& \sum_{t=1}^n I(\tilde{\M}_2\tilde{X}^{t-1}\tilde{Y}^{t-1}\tilde{Z}^{t-1};\tilde{Z}_t) \nonumber \\
	&& \qquad + I(\tilde{\M}_1\tilde{X}^{t-1}\tilde{Y}^{t-1};\tilde{Y}_t)\IEEEeqnarraynumspace\\
	&=& \sum_{t=1}^n I(\tilde{\M}_2\tilde{X}^{t-1}\tilde{Y}^{t-1};\tilde{Z}_t) + I(\tilde{\M}_1\tilde{X}^{t-1}\tilde{Y}^{t-1};\tilde{Y}_t)\IEEEeqnarraynumspace\label{eq:divergence_markovcahins}\\
	&=& \sum_{t=1}^n I(\tilde{U}_{2,t};\tilde{Z}_t) + I(\tilde{U}_{1,t};\tilde{Y}_t)\label{eq:divergence_end1}\\
	&=& n[I(\tilde{U}_{2,T};\tilde{Z}_T|T) + I(\tilde{U}_{1,T};\tilde{Y}_T|T)]\\
	&\leq& n[I(\tilde{U}_{2,T}T;\tilde{Z}_T) + I(\tilde{U}_{1,T}T;\tilde{Y}_T)]\\
	&=& n [I(\tilde{U}_2;\tilde{Z}) + I(\tilde{U}_1;\tilde{Y})]\label{theta_ub2}.
\end{IEEEeqnarray}
Here \eqref{eq:dp_ineq_relative_entropy} is obtained by the data processing inequality for Kullback-Leibler divergence; \eqref{eq:divergence_chainrule} by the chain rule; \eqref{eq:divergence_markovcahins} by the Markov chain $\tilde{Z}^{t-1} \to (\tilde{Y}^{t-1},\tilde{X}^{t-1}) \to \tilde{Z}_t$; and \eqref{eq:divergence_end1}--\eqref{theta_ub2} by the definitions of $\tilde{U}_{1,t},\tilde{U}_{2,t},\tilde{U}_1,\tilde{U}_2,\tilde{Y}$ in \eqref{eq:Utilde_t} and \eqref{eq:Utilde} and by defining $\tilde{Z} = \tilde{Z}_T$ where $T$ is uniform over $\{1,\ldots,n\}$ independent of all other random variables.

Observe the Markov chain $\tilde{U}_{2,t} \to \tilde{Y}_{t} \to \tilde{Z}_t$ for any $t$, and thus $\tilde{U}_2 \to \tilde{Y}\to \tilde{Z}$ holds by construction for any $n$.

The second desired Markov chain $\tilde{U}_1 \to \tilde{X} \to \tilde{Y}$ holds only in the limit as $n \to \infty$.
To see this, notice that $\tilde{\M}_1 \to \tilde{X}^n \to \tilde{Y}^n$ forms a Markov chain, and thus similar to the analysis in \cite[Section V.C]{HWS20}:
\begin{IEEEeqnarray}{rCl}
	0 &=& I(\tilde{\M}_1;\tilde{Y}^n|\tilde{X}^n) \\ 
	&\geq& H(\tilde{Y}^n|\tilde{X}^n)  - H(\tilde{Y}^n|\tilde{X}^n\tilde{\M}_1) \nonumber \\
	&& + D(P_{\tilde{X}^n\tilde{Y}^n}||P_{XY}^n) + \log{\Delta_{n}}
	\label{MC1proofstep1}\\
	&{\geq}& n[H(\tilde{Y}_T|\tilde{X}_{T}) + D(P_{\tilde{X}_{T}\tilde{Y}_T}||P_{XY})] +\log{\Delta_{n}} \nonumber\\
	&& - H(\tilde{Y}^n|\tilde{X}^n\tilde{\M}_1) \label{MC1proofstep2}\\
	&\geq& n[H(\tilde{Y}_T|\tilde{X}_{T}) + D(P_{\tilde{X}_{T}\tilde{Y}_T}||P_{XY})] +\log{\Delta_{n}} \nonumber \\ 	&&-  \sum_{t=1}^{n}H(\tilde{Y}_t|\tilde{X}_{t}\tilde{X}^{t-1}\tilde{Y}^{t-1}\tilde{\M}_1)\label{MC1proofstep3}\\
	&=& n[H(\tilde{Y}_T|\tilde{X}_{T}) + D(P_{\tilde{X}_{T}\tilde{Y}_T}||P_{XY})] +\log{\Delta_{n}} \nonumber \\
	&&-  \sum_{t=1}^{n}H(\tilde{Y}_t|\tilde{X}_{t}\tilde{U}_{1,t})\label{MC1proofstep4}\\
	&\geq& n[H(\tilde{Y}_T|\tilde{X}_{T}) - H(\tilde{Y}_T|\tilde{X}_T,\tilde{U}_{1,T},T) ]+ \log{\Delta_{n}}  \IEEEeqnarraynumspace\label{MC1proofstep4b}\\
	&\geq& nI(\tilde{Y};\tilde{U}_1|\tilde{X}) + \log{\Delta_{n}},\label{MC1proofstep5}
\end{IEEEeqnarray}
where \eqref{MC1proofstep2} holds by the super-additivity property in \cite[Proposition 1]{tyagi2019strong}; \eqref{MC1proofstep3} by the chain rule and since conditioning reduces entropy; \eqref{MC1proofstep4} by the definition of $\tilde{U}_{1,t}$ in \eqref{eq:Utilde_t}; \eqref{MC1proofstep4b} by the non-negativity of the Kullback-Leibler divergence, and by recalling that $T$ is uniform over $\{1,\ldots,n\}$ independent of all other random quantities, and finally \eqref{MC1proofstep5} holds by the definitions of $\tilde{U}_1,\tilde{X}, \tilde{Y}$ in \eqref{eq:Utilde}.

To sum up, we have proved so far in \eqref{R1lb}, \eqref{R2lb}, \eqref{theta_ub}, \eqref{theta_ub2}, and \eqref{MC1proofstep5}, that for all $n\geq 1$ there exists a joint pmf $P_{\tilde{X}\tilde{Y}\tilde{Z}\tilde{U}_1\tilde{U}_2}^{(n)}$ (abbreviated as $P^{(n)}$) so that the following conditions hold (where $I_{P^{(n)}}$ indicates that the mutual information should be calculated according to the pmf $P^{(n)}$):
\begin{subequations}\label{eq:conditions}
	\begin{IEEEeqnarray}{rCl}			
		R_1 &\geq  &\big(I_{P^{(n)}}(\tilde{U}_1;\tilde{X}) + g_1(n)\big)\cdot g_2(n,\eta) ,\label{eq:R111} \IEEEeqnarraynumspace \\
		R_2 &\geq  &\big(I_{P^{(n)}}(\tilde{U}_2;\tilde{Y}) + g_1(n)\big)\cdot g_2'(n,\eta), \label{eq:R222}\\
		\theta_2 &\leq &\big(I_{P^{(n)}}(\tilde{U}_2;\tilde{Z}) + I_{P^{(n)}}(U_1';\tilde{Y})\big)g_3(n) + g_4(n) \label{eq:theta222}\IEEEeqnarraynumspace\\
		g_5(n) &\geq& I_{P^{(n)}}(\tilde{Y};\tilde{U}_1|\tilde{X}) , \label{eq:last_cond}
	\end{IEEEeqnarray}
\end{subequations}
for some functions $g_1(n), g_2(n,\eta), g_2'(n,\eta), g_3(n), g_4(n), g_5(n)$ with the following asymptotic behaviors:
\begin{IEEEeqnarray}{rCl}
\lim_{n\to \infty} g_1(n) = \lim_{n\to \infty} g_4(n)=\lim_{n\to \infty} g_5(n)& = & 0\\
\lim_{n\to \infty} g_3(n) &= & 1\\
\lim_{n\to \infty} g_2(n,\eta)=\lim_{n\to \infty} g_2'(n,\eta) & =  &\frac{1-\epsilon-\eta}{1-\eta}. \IEEEeqnarraynumspace
\end{IEEEeqnarray}
By the  Markov chains $\tilde{X} \to \tilde{Y} \to \tilde{Z}$ and $\tilde{U}_2 \to \tilde{Y} \to \tilde{Z}$ we can further conclude that 
	\begin{IEEEeqnarray}{rCl}	
		P_{\tilde{X}\tilde{Y}\tilde{Z}\tilde{U}_1\tilde{U}_2}^{(n)} &=& P_{\tilde{X}\tilde{Y}\tilde{Z}}^{(n)}\cdot P_{\tilde{U}_1\tilde{U}_2|\tilde{X}\tilde{Y}}^{(n)},\IEEEeqnarraynumspace \label{eq:pmf_n}
	\end{IEEEeqnarray}	
		
	The proof in this section  is  concluded by letting $n\to \infty$ and $\eta\downarrow 0$, and noting that by \eqref{eq:last_cond} the limiting pmf of the sequence $P^{(n)}$ satisfies the Markov condition $\tilde{U}_1\to \tilde{X}\to \tilde{Y}$. More precisely, we first observe that by  Carath\'eodory's theorem \cite[Appendix C]{ElGamal} for each $n$ there must exist random variables  $\tilde{U}_1$ and $\tilde{U}_2$ satisfying \eqref{eq:conditions} and \eqref{eq:pmf_n} over  alphabets of sizes\vspace{-0.05cm}
\begin{align}
	\vert \tilde{\mathcal{U}}_1\vert &\leq \vert \mathcal{X}\vert\cdot\vert \mathcal{Y}\vert + 2,\\
	\vert \tilde{\mathcal{U}}_2 \vert &\leq \vert \tilde{\mathcal{U}}_1 \vert\cdot|\mathcal{X}|\cdot\vert \mathcal{Y}\vert + 1.
\end{align}
Then we invoke the Bolzano-Weierstrass theorem and  consider a sub-sequence  $P_{\tilde{X}\tilde{Y}\tilde{Z}\tilde{U}_1\tilde{U}_2}^{(n_k)}$ that converges to a limiting pmf $P_{XYZU_1''U_2}^{*}$. For this limiting pmf, which we abbreviate by $P^*$, we conclude by \eqref{eq:R111}--\eqref{eq:theta222}:
	\begin{IEEEeqnarray}{rCl}			
		R_1& \geq &(1-\epsilon)I_{P^{*}}({U}_1'';{X})\label{eq:R_1_f} \\
		R_2 &\geq  &(1-\epsilon)I_{P^{*}}({U}_2;{Y}) \\
		\theta_2 &\leq &I_{P^{*}}({U}_2;{Z}) + I_{P^{*}}(U_1'';{Y}).\label{theta_2_f}\end{IEEEeqnarray}
Notice further that since for any $k$ the pair $(\tilde{X}^{n_k},\tilde{Y}^{n_k})$ lies in the jointly typical set $\mathcal{T}^{(n_k)}_{\mu_{n_k}}(P_{XY})$, we have  $\vert P_{\tilde{X}\tilde{Y}} - P_{XY}\vert \leq \mu_{n_k}$ and thus the limiting pmf $P^*$ satisfies $P^*_{XY}=P_{XY}$. Moreover, since for each $n_k$ the random variable $\tilde{Z}$ is drawn according to $P_{Z|Y}$ given $\tilde{Y}$, irrespective of $\tilde{X}$, the limiting pmf also satisfies $P_{Z|XY}^*=P_{Z|Y}$. 
We also notice that   under $P^*$ the Markov chain
\begin{IEEEeqnarray}{rCl}\label{eq:MC2_1}
U_2\to Y \to Z,
\end{IEEEeqnarray}	
holds because  $\tilde{U}_2\to \tilde{Y} \to \tilde{Z}$ also forms a Markov chain for any $n_k$. Finally, by continuity considerations and by \eqref{eq:last_cond}, the following Markov chain must hold under $P^*$:
\begin{IEEEeqnarray}{rCl}\label{eq:MC2_2}
U_1'' \to X \to Y
\end{IEEEeqnarray}

To summarize, we establish the existence of a pmf $P^*_{XYZU_1''U_2}$ with $P_{XYZ}^*=P_{XY}P_{Z|Y}$, and satisfying the Markov chains \eqref{eq:MC2_1}--\eqref{eq:MC2_2} and the constraints \eqref{eq:R_1_f}--\eqref{theta_2_f}.

\subsection{Combining Constraints from Decisions at {\Rel} and \Rec}\label{sec:part3}
The previous Subsections~\ref{sec:part1}--\ref{sec:part2} established the existence of random variables $U_1'$, $U_1''$, and $U_2$ satisfying the three Markov chains $U_1'\to X \to Y$,  $U_1'' \to X \to Y$, and $U_2 \to Y \to Z$, and constraints
\begin{IEEEeqnarray}{rCl}\label{eq:R1R2thetaRthetaD}
	R_1& \geq &(1-\epsilon) \max \left\{I(U_1';X) ; I(U_1'';X)\right\}, \\
	R_2& \geq & (1-\epsilon)I(U_2;Y), \\
	\theta_1 &\leq& I({U_1'};{Y}),\\
	\theta_2 &\leq& I(U_2;{Z}) + I(U_1'';{Y}).\label{eq:R1R2thetaslasteq}
\end{IEEEeqnarray}

The proof is concluded by showing that for each choice of $U_1', U_1'', U_2$, constraints \eqref{eq:R1R2thetaRthetaD}--\eqref{eq:R1R2thetaslasteq} are relaxed if one replaces both $U_1'$ and $U_1''$ with the same suitably chosen random variable $U_1$. In fact, we choose $U_1=U_1''$ if $I({U_1'};Y) \leq I(U_1'';Y)$ and  we choose $U_1={U_1'}$ otherwise. For this choice, \eqref{eq:R1R2thetaRthetaD}--\eqref{eq:R1R2thetaslasteq} imply 
\begin{IEEEeqnarray}{rCl}
	R_1& \geq & (1-\epsilon) I(U_1;X) , \\
	R_2& \geq & (1-\epsilon)I(U_2;Y), \\
	\theta_1 &\leq& I(U_1;Y) ,\\
	\theta_2 &\leq& I(U_2;{Z}) + I(U_1;{Y}).
\end{IEEEeqnarray}
Since the Markov chains $U_1 \to X \to Y$ and $U_2 \to Y \to Z$ hold by definition, this concludes our  converse proof for the result in \eqref{eq:E1}.
\end{document}